\documentclass[12pt, a4paper, twoside]{article}
\usepackage[a4paper, left=3cm,right=2cm]{geometry}
\usepackage{hyperref}
\usepackage{amsfonts}
\usepackage{amsmath}
\usepackage{setspace}
\usepackage{color}

\usepackage[utf8,applemac]{inputenc}
\usepackage{tensor}
\usepackage{cite}
\usepackage{tikz}
\usepackage{graphicx}
\usepackage[font=small, font+=it, format=hang, margin={1cm, 1cm}]{caption}
\graphicspath{{figure/}}

\usepackage{dcolumn}
\usepackage{bm}

\newcommand{\bea}{\begin{eqnarray}}
\newcommand{\eea}{\end{eqnarray}}
\newcommand{\be}{\begin{equation}}
\newcommand{\ee}{\end{equation}}
\def\nn{\nonumber}

\def\p{\partial}

  \newcommand{\beqs}{\begin{eqnarray}}
\newcommand{\eeqs}{\end{eqnarray}}

\title{Vacua of the gravitational field}

\begin{document}


\setcounter{tocdepth}{2}

\begin{titlepage}

\begin{flushright}\vspace{-3cm}
{\small
\today }\end{flushright}
\vspace{0.5cm}

\begin{center}
{{ \LARGE{\bf{Vacua of the gravitational field\\ }}}} \vspace{5mm}

\centerline{\large{\bf{Geoffrey Comp\`{e}re\footnote{e-mail: gcompere@ulb.ac.be}, Jiang Long\footnote{e-mail:
Jiang.Long@ulb.ac.be}}}}

\vspace{2mm}
\normalsize
\bigskip\medskip
\textit{Universit\'{e} Libre de Bruxelles and International Solvay Institutes\\
CP 231, B-1050 Brussels, Belgium
}

\vspace{25mm}

\begin{abstract}
\noindent
{The Poincar\'e invariant vacuum is not unique in quantum gravity. The BMS supertranslation symmetry originally defined at null infinity is spontaneously broken and results in inequivalent Poincar\'e vacua. In this paper we construct the unique vacua which interpolate between past and future null infinity in BMS gauge and which are entirely characterized by an arbitary Goldstone boson defined on the sphere which breaks BMS invariance. We show that these vacua contain a defect which carries no Poincar\'e charges but which generically carries superrotation charges. We argue that there is a huge degeneracy of vacua with multiple defects. We also present the single defect vacua with its canonically conjugated source which can be constructed from a Liouville boson on the stereographic plane. We show that positivity of the energy forces the stress-tensor of the boson to vanish as a boundary condition. Finite superrotations, which turn on the sources, are therefore physically ruled out as canonical transformations around the vacua. Yet, infinitesimal superrotations are external symplectic symmetries which are associated with conserved charges which characterize the Goldstone boson. 
}

\end{abstract}


\end{center}

\end{titlepage}

\newpage
\tableofcontents

\section{Introduction}

The structure of asymptotically flat spacetimes is a classical topic in general relativity which has been studied for more than 50 years. One highlight is the existence of a symmetry group at future null infinity
which extends the Poincar\'e group, the so-called BMS group from the name of its founders Bondi, van der Burg, Metzner and Sachs \cite{Bondi:1962px,Sachs:1962wk}. The BMS group includes the familiar rotations and Lorentz boosts but extends the translation abelian normal subgroup into a larger abelian normal subgroup: the supertranslations. (For a review of the asymptotic structure of the gravitational field  at future null infinity and spatial infinity, see e.g. \cite{ABR}). 

Recently, it was understood that the symmetry group of asymptotically flat spacetimes should be the diagonal subgroup of the direct product of the symmetry group at both future and past null infinity: there is only one BMS algebra and the S-matrix is BMS invariant \cite{Strominger:2013jfa}. It was further understood that the vacuum spontaneously break the BMS supertranslation symmetry which leads to the existence of a Goldstone boson which labels inequivalent vacua \cite{He:2014laa}. Also, two new connections emerged: the Ward identities of the BMS symmetry naturally account for the leading tree-level soft gravitons theorems \cite{He:2014laa}. Moreover, supertranslations can also be related to physical displacements known as memory effects \cite{Strominger:2014pwa}.

Yet, all the consequences of these structures have been established so far only in the asymptotic regions either at future or past null infinity. The main aim of this paper is to infer what these structures tell us about the bulk spacetime. Let us imagine by analogy a world where theoretical physicists only knew about the asymptotic structure of $AdS_3$ spacetimes without any information about the bulk $AdS_3$ physics. In this fiction, these physicists would only know that the asymptotic symmetry group is the double copy of the Virasoro group \cite{Brown:1986nw} and they would know the Fefferman-Graham asymptotic expansion \cite{Fefferman:1985aa}. Since the symmetry generators admit a finite asymptotic expansion for Einstein gravity, one can easily exponentiate them in order to obtain an interesting phase space: the $AdS_3$ spacetime equipped with ``holographic gravitons'' \cite{Banados:1998gg,Rooman:2000ei} (for recent views on this topic, see e.g. \cite{Barnich:2010eb,Barnich:2012rz,Barnich:2013yka,Garbarz:2014kaa,Barnich:2014kra,Barnich:2015uva,Compere:2015knw})\footnote{Note that it would then be straightforward to add the zero modes to the functions appearing in the resulting metric as a solution generating technique. This would result in a derivation of the metric of the BTZ black holes \cite{Banados:1992wn} and the conical angle deficits in $AdS_3$ \cite{Deser:1984dr} (still equipped with holographic gravitons \cite{Banados:1998gg}).}. In our view, we have a similar situation for four-dimensional asymptotically flat spacetimes. The symmetry group is defined at null infinity but these symmetries have not yet been exponentiated in the bulk in order to generate the corresponding phase space.

There are two obstacles to this program. The first one is technical: the asymptotic series of the BMS supertranslation generator in the radial asymptotic expansion is infinite and depends upon an arbitrary function on the unit sphere. We were however able to overcome this difficulty in the vacuum case by explicitly resumming the radial asymptotic expansion presented in detail in \cite{Barnich:2010eb,Barnich:2011mi}. The second one is conceptual: aren't the Goldstone boson and the asymptotic symmetries defined in the asymptotic region only? In fact, the answer is negative. At least as an asymptotic series at future null infinity, the BMS asymptotic symmetry holds at each order after defining the suitable bracket of infinitesimal symmetries \cite{Barnich:2010eb,Barnich:2013oba}. Now, in a  restricted phase space where no local degree of freedom is present, the symmetries also lead to conserved charges which represent the symmetry algebra everywhere in the bulk spacetime \cite{Compere:2014cna}. Such symmetries  were dubbed symplectic symmetries and further characterized in \cite{Compere:2015knw,Compere:2015bca}. In the phase space of vacua, the BMS supertranslations are symplectic symmetries as well, as we will discuss in detail, but their associated charges will be found to be identically zero. This result dovetails with a similar observation made recently in the closely related context of stationary spacetimes  \cite{Flanagan:2015pxa} and black holes \cite{Hawking:2016msc}. 

The main remaining issue that we will address is the following: if there is a non-trivial Goldstone boson, which conserved charges characterize it? Part of the answer comes from the suggestion to consider an extension of the BMS group  \cite{Barnich:2009se}\footnote{See also footnote 18 in \cite{Banks:2003vp}.}: Lorentz boosts and rotations form together the group of global conformal transformations of the conformal sphere at future null infinity. Local conformal transformations, associated with superrotations, are then a natural extension of the symmetry algebra which leads to interesting algebraic structures \cite{Barnich:2010eb,Barnich:2011mi}. These symmetry structures have also been claimed to be related to the subleading soft graviton theorem \cite{Kapec:2014opa} in turn related to the spin memory effect \cite{Pasterski:2015tva}. In this paper, we will argue that finite superrotations cannot be considered as physical canonical transformations for at least three reasons: they would lead to a violation of positivity of the energy, they would lead to symplectic flux at future null infinity and they would forbid the existence of a variational principle for the vacua. However, we will show that infinitesimal superrotations are \emph{external} symplectic symmetries of the vacua, i.e. symplectic symmetries which are not canonical transformations\footnote{A possible relationship with the recent work \cite{Troessaert:2015nia} remains to be understood.}. Yet, the superrotations are associated to physical conserved charges which can then be used to characterize the Goldstone boson of the vacua.

The rest of the paper is organized as follows. In Section \ref{prop} we present the metric of the vacua obtained by exponentiating the infinitesimal BMS supertranslations and we discuss its properties. In Section \ref{vws} we present the metric obtained by exponentiating both the BMS supertranslations and superrotations. We argue that the metric should not be considered as physical. We interpret it as containing the source conjugated to the Goldstone boson. In Section \ref{symplectic} we discuss the existence of symplectic symmetries of the vacua. In Section \ref{multi} we generalize the construction of vacua to the multi-centered case and we conclude in Section \ref{ccl}. 

\section{Vacua}
\label{prop}

We define the \emph{global Poincar\'e vacuum} as the Minkowski vacuum
\bea
ds^2 = -du^2-2 du dr + 2 r^2 \gamma_{z \bar z} dz d\bar z,\qquad  \gamma_{z \bar z} = \frac{2}{(1+z \bar z)^2}.\label{Mink}
\eea
We used retarded time $u=t-r$ and stereographic coordinates defined as $z = e^{i\varphi} \cot \frac{\theta}{2}$, $\bar  z = e^{-i\varphi} \cot \frac{\theta}{2}$.\footnote{For future reference, we denote the unit metric on the round sphere as $\gamma_{AB}$, $A,B=z,\bar z$ and its determinant as $\gamma$. It has all Christoffel symbols vanishing except $\Gamma^z_{zz} =-2 \p_z \phi$, $\Gamma^{\bar z}_{\bar z \bar z} =-2 \p_{\bar z} \phi$ where we define $e^{-2\phi} \equiv \gamma_{z \bar z}$. We will denote  its covariant derivative as $D_A$. Its Ricci tensor and scalar obey $R_{AB}=\gamma_{AB}$, $R = 2$. Note that the north pole is at $z,\bar z \rightarrow \infty$ and the south pole is at $z=\bar z = 0$.}

The existence of other inequivalent vacua is signaled by the existence of BMS supertranslations at future null infinity which constitute together with the Lorentz group the set of asymptotic symmetries of gravity with zero cosmological constant. The generator of supertranslations reads in BMS gauge around the global Poincar\'e vacuum as 
\be
\chi_T =T \partial_u+(\frac{1}{2} D^A D_A T+\mathcal{O}(\frac{1}{r}))\partial_r+(-\frac{1}{r} {D}^AT+\mathcal{O}(\frac{1}{r^2}))\partial_A\label{ST}
\ee
where $T=T(z,\bar z)$ is a regular real function on the sphere. Acting with the Lie derivative of this vector field on the Minkowski vacuum, one turns on a phase space variable $C$ which is defined in the asymptotic region and which transforms under supertranslations as
\bea
\delta_T C= T(z,\bar z). \label{BMST}
\eea

We define the vacua as the zero-energy solutions of Einstein gravity with asymptotically flat boundary conditions which break BMS supertranslation invariance at future null infinity. Such solutions are characterized by the Goldstone boson of spontaneously broken BMS supertranslation invariance $C$. We assume in this section that it only depends upon the sphere coordinates and we denote it by $C(z,\bar z)$. We will go back to this assumption in Section \ref{multi}. 

\subsection{Metric}

The metric of the vacua in the BMS coordinate system at future null infinity can be deduced from the form of the generator \eqref{ST}. In principle, one needs to exponentiate this generator in order to obtain a finite diffeomorphism. This finite diffeomorphism will depend upon the exponentiation of $T(z,\bar z)$ which we define as $C(z,\bar z)$. One then applies this finite diffeomorphism to the global Poincar\'e vacuum to obtain the other vacua. This diffeomorphism is a solution generating diffeomorphism which sends a solution to another inequivalent solution, as we will extensively describe in the following. 

The technical difficulty in this program is the fact that the supertranslation generator is only defined in the literature as an asymptotic series expansion close to future null infinity. In an analogue toy model in $3$ dimensional gravity the asymptotic symmetries admit a finite radial expansion and the exponentiation can be performed exactly, see  \cite{Rooman:2000ei} for the $AdS_3$ case and \cite{Barnich:2015uva} for the flat case. Here, we shall not discuss all harsh technical steps required in 4 dimensions, but we shall present the final result which can be written in a simple closed form. The details of the solution generating diffeomorphism can be found in the Appendix of \cite{Compere:2016hzt}.

The metric of the vacua can be written exactly as
\bea
ds^2 =\left[\begin{matrix}
-1&-\frac{r}{\sqrt{r^2+U}}&\frac{1}{2}D^BC_{AB}-D_A\sqrt{r^2+U}&\\-\frac{r}{\sqrt{r^2+U}}&0&0&\\\frac{1}{2}D^BC_{AB}-D_A\sqrt{r^2+U} &0&(r^2+2U)\gamma_{AB}+C_{AB}\sqrt{r^2+U}&\\
\end{matrix}\right] \label{vacua2}
\eea
with $C_{AB}=-(2D_AD_B-\gamma_{AB} D^2)C$ and $U=\frac{1}{8}C_{AB}C^{AB}$. The metric depends upon a single function $C(z,\bar z)$ defined on the sphere. When $C(z,\bar z) = 0$ the metric is exactly the global Poincar\'e vacuum \eqref{Mink}.

In $z,\bar z$ notation, one can write it as 
 \be
 ds^2=\left[\begin{matrix}
-1&-\frac{1}{\sqrt{1+\frac{U}{r^2}}}&\frac{1}{2}  D^zC_{zz}-\partial_z\sqrt{r^2+U} &\frac{1}{2} D^{\bar z} C_{\bar z \bar z}-{\partial}_{\bar z}\sqrt{r^2+U}&\\
-\frac{1}{\sqrt{1+\frac{U}{r^2}}}&0&0&0&\\
\frac{1}{2} D^z C_{zz}-\partial_z \sqrt{r^2+U} &0&C_{zz}\sqrt{r^2+U}& \gamma_{z \bar z}(r^2+2U)&\\
\frac{1}{2}D^{\bar z} C_{\bar z \bar z}-{\partial}_{\bar z} \sqrt{r^2+U}&0& \gamma_{z \bar z}(r^2+2U)&C_{\bar z \bar z}\sqrt{r^2+U}&\\
\end{matrix}\right] \label{vacua}
\ee
where indices are raised with $\gamma^{z \bar z}$ and
\bea
U&\equiv&\frac{1}{4}C_{zz}C^{zz} = \frac{1}{4}C_{\bar z \bar z}C^{\bar z \bar z}, \label{defU} \\
 C_{zz}&\equiv& -2 D_z \p_z C = - 2(\partial_z \partial_z C+2\partial_z C \partial_z \phi ), \nn\\
    C_{\bar z \bar z}& \equiv&- 2 D_{\bar z} \p_{\bar z} C  = -2({\partial}_{\bar z} {\partial}_{\bar z} C+2 {\partial}_{\bar z} C {\partial}_{\bar z}\phi ).
\eea
Note the equalities $D^z C_{zz} = {\bar \partial}^z C_{zz} \equiv \gamma^{z \bar z} { \partial}_{\bar z}  C_{zz} $ and $D^{\bar z} C_{\bar z \bar z} = {\partial}^{\bar z}C_{\bar z \bar z}  \equiv \gamma^{\bar z z}\partial_zC_{\bar z \bar z} $.

By construction, the metric is asymptotically flat at future null infinity and it falls into the BMS gauge. The metric is also asymptotically flat at past null infinity when switched to  BMS$_-$ gauge. An explicit match with standard asymptotic expansions is provided in Appendix \ref{BMSg}.

\subsection{Properties}

The metric is defined in a coordinate system which breaks down at $r=0$. This locus is of crucial interest. 
The coordinate singularity at $r=0$ differs from the familiar one in spherical coordinates. In the global vacuum, the sphere at constant radius $r$ and constant $u$ shrinks upon decreasing $r$ to a space point at $r=0$ which is the origin of Euclidean space. Instead, in the metric \eqref{vacua} the induced metric $g_{AB}$ at constant $u,r$ only admits a vanishing determinant at $r=0$ but the induced metric does not entirely vanish. Indeed, after using  \eqref{defU}
we find that
\bea
\text{Det}\left( 2U \gamma_{AB} + \sqrt{U}  C_{AB}  \right) = 0. 
\eea
We will prove in Section \ref{charges} that spheres at constant $r,u$ cannot be shrunk to zero size since non-vanishing conserved charges can be defined on such spheres, otherwise the charges would be identically zero. We will however need an entire machinery to prove it. It implies that there is a defect of some nature at $r=0$ at constant $u$ or beyond that locus. There is no global conical deficit or angular defect since the metric is smooth for $r > 0$.

Since the metric is obtained by a finite coordinate change from Minkowski, it is Riemann flat everywhere, except possibly at singular points of the coordinate change, which would lead to a defect. In general, the metric is stationary but nonaxisymmetric. 

The metric depends upon $C(z,\bar z) $ only indirectly, through $C_{zz}$ and $C_{\bar z\bar z}$. This is related to the fact that the 4 translations of Minkowski spacetime \eqref{ST} where 
$T(z,\bar z)$ is a real linear combination of 
\bea
1,\;\frac{1}{1+z \bar z},\; \frac{z}{1+z \bar z},\; \frac{\bar z}{1+z \bar z}
\eea
are Killing symmetries. Therefore, they do not generate any Goldstone boson $C$. These 4 functions are exactly the basis of the zero modes of the differential operator $2D_A D_B - \gamma_{AB}D^2$. These are the lowest spherical harmonics $l=m=0$ and $l=1$, $m=-1,0,1$. Equivalently, these four zero modes are the unique zero mode of $D^2$ (time translations) and the three zero modes of $D^2+2$ (space translations).

Since the vacua are just Minkowski vacuum written in different coordinate systems, the metric \eqref{vacua} admits 10 Killing vectors. These Killing vectors however have components $\xi^\mu$ which depend upon the field $C(z,\bar z)$ since they are obtained as $\xi^\mu = \frac{\p x^\mu}{\p x_s^\mu} \xi^\mu_s$ where $\xi^\mu_s$ are the standard components of the Killing vectors in global coordinates $x_s^\mu$ \eqref{Mink}. Given the unusual form of these Killing vectors, we find no contradiction to reconcile translation invariance and the presence of a defect.

After using the following identity
\be
D^BC_{AB}=-D_{A}(D^2+2)C,\label{propC}
\ee
we can write the metric for the vacua as
\bea
ds^2&=&-du^2-2dud(\sqrt{r^2+U}  + \frac{1}{2}(D^2+2)C)+ \left( (r^2 + 2U)\gamma_{AB} + \sqrt{r^2+U}  C_{AB} \right)dz^Adz^B. \nonumber\\
&=& -du^2-2dud\rho+ g_{AB}(\rho,z^C) dz^Adz^B.
\eea
Here, we need to be careful in the definition of the radial coordinate $\rho$. Out of the 4 arbitrary first spherical harmonics of $C$, 3 are annihilated by $D^2+2$ but the lowest constant harmonic $C_{(0,0)}$ is not. Hence, in the last step, we define the new radial coordinate $\rho$ as
\bea
\rho = \sqrt{r^2+U}  + \frac{1}{2}(D^2+2)C -C_{(0,0)} = r +O(r^0).\label{defr}
\eea
It agrees asymptotically with $r$ in retarded coordinates and it does not depend upon the four arbitrary first harmonics of $C$. 
Static coordinates are obtained by defining $t = u + \rho$. The metric reads
\bea
ds^2&=&-dt^2 + d\rho^2 + g_{AB}(\rho,z^C) dz^Adz^B, \nonumber\\
g_{AB} &=& (\rho-C+C_{(0,0)})^2 \gamma_{AB} -2(\rho - C+C_{(0,0)}) D_A D_B C + D_A D_E C D_B D^E C.\label{m1}
\eea
The definition \eqref{defr} ensures that time translations only shift $t$ or $u$ but not $\rho$. Quite surprizingly, the metric is very simple in static gauge: it is quadratic in both $C$ and $\rho$.

Since we did not impose any parity conditions on the field $C$, the metric \eqref{m1} provides an explicit example of a Ricci flat solution with finite charges (as we will show in Section \ref{charges}) which explicitly breaks the parity conditions at spatial infinity which are usually imposed \cite{Regge:1974zd}.

The metric in advanced coordinates $(v,r,z,\bar z)$ is simply obtained after the change of coordinates $v = t+\rho=u+2 \rho$. We use the same coordinates $z^A$ at $\mathfrak{I}^-$ as in the bulk spacetime and at $\mathfrak{I}^+$: a light ray starting at $(z,\bar z)$ at $\mathfrak{I}^-$ will reach the antipodal point $(-\frac{1}{\bar z},-\frac{1}{z})$ at $\mathfrak{I}^+$\footnote{Note that the metric and Laplacian on the unit sphere are invariant under the antipodal map.}. We find that the metric in advanced coordinates $x^\mu_{(-)} = \{ v,\rho,z^A \}$ reads as
\bea
ds^2&=&-dv^2 + 2 \p_r(\sqrt{r^2+U}) dv dr  + 2 \p_{A} ( \sqrt{r^2+U}  + \frac{1}{2}(D^2+2)C ) dv dz^A \nonumber \\
&& + \left( (r^2 + 2U)\gamma_{AB}+ \sqrt{r^2+U}  C_{AB} \right)dz^Adz^B\label{madv}
\eea
where $U = \frac{1}{8}C_{AB}C^{AB}$ and $C_{AB} = -(2 D_A D_B -\gamma_{AB}D^2)C(z,\bar z)$. The metric \eqref{madv} is explicitly asymptotically flat in BMS gauge around past null infinity as detailed in Appendix \ref{BMSg}. We have therefore found the explicit map between future null infinity and past null infinity thanks to the knowledge of the metric everywhere in the bulk. The Goldstone boson of spontaneously broken supertranslation invariance at future and past null infinity are related as
\bea
C(z,\bar z) = C^{(-)}(z,\bar z).\label{match}
\eea
where $C^{(-)}$ is defined in \eqref{defCm}, up to the four first harmonics which are left unfixed. 

Therefore, the vacua do not obey in general the antipodal matching conditions
\bea
C(z,\bar z) = -C^{(-)}(-\frac{1}{\bar z},- \frac{1}{z}).\label{anti}
\eea
as stated in (3.2) of \cite{He:2014laa} or as stated in (2.12) of \cite{Pasterski:2015tva} (for comparison, note their different sign convention for $C^{(-)}$). 
Note that the antipodal matching conditions \eqref{anti} trivially hold as a consequence of \eqref{match} when $C$ is restricted to be parity odd. 

In advanced coordinates, the generator of supertranslations reads as 
\be
\chi^{(-)}_T =T^{(-)}(z^A) \partial_v+(- \frac{1}{2} D^2 T^{(-)}+\mathcal{O}(\frac{1}{r}))\partial_r+( \frac{1}{r} D^AT^{(-)}+\mathcal{O}(\frac{1}{r^2}))\partial_A.\label{STm}
\ee
Infinitesimal coordinate transformations generated by $\chi^{(-)}$ preserve BMS gauge at past null infinity. The transformation law for $\delta C_{AB}$ allows to infer the transformation law for $C^{(-)}$ as $\delta_{T^{(-)}} C^{(-)} = - T^{(-)}$. However, for the 4 translations, $\delta C_{AB} = 0$ and the transformation law of $C^{(-)} $ is not fixed by this argument.

A subtle point is the relationship between the vector fields $\chi_T$ and $\chi_T^{(-)}$. One can check using the coordinate changes that they are not equal. This is related to the fact that the change of coordinates from advanced to retarded coordinates depends upon the very field $C$ which is varied by the action of supertranslations. The identification of the supertranslations between future and past null infinity is
\bea
\chi_T = \chi^{(-)}_T - \delta_{\chi_T} x_{(-)}^\mu \frac{\p}{\p x_{(-)}^\mu} \qquad \text{or} \qquad \chi^{(-)}_T = \chi_T - \delta_{\chi_{T^{(-)}}} x_{(+)}^\mu \frac{\p}{\p x_{(+)}^\mu} \quad \label{matchchi}
\eea
where the variation acts on the field $C$ as $\delta_{\chi_T} C(z,\bar z) =T(z,\bar z)$. For usual translations, one has $\delta_{\chi_T} x_{(-)}^\mu = 0$ and there is a unique vector field $\chi_T = \chi_T^{(-)}$. For generic supertranslations, we find explicitly 
\bea
\chi_T^{(-)v} &=& \frac{\p v}{\p x^\mu_{(+)}} \chi_T^\mu - \delta_{\chi_T} v = \chi_T^u + 2 \chi_T^r - 2 \delta_{\chi_T} \rho \nonumber \\
&=& T(z^A) + D^2 T(z^A) - (D^2+2) T(z^A)+ 2 T_{(0,0)}  = - T(z^A)+ 2 T_{(0,0)} . \label{law1}
\eea
Therefore, the identification \eqref{matchchi} leads to the following identification between BMS supertranslations at future and null infinity (except the time translation)
\bea
T^{(-)}(z,\bar z) = - T(z,\bar z).\label{TTm}
\eea
The only exception is the time translation which is simply identified as $\p_u = \p_v$ ($T^{(-)} = T=1$). One can check that the relationship is obeyed in the case of exact translations. The parity-odd BMS supertranslations and time translations obey the antipodal matching conditions
\bea
T(z,\bar z) = T^{(-)}(-\frac{1}{\bar z}, -\frac{1}{z}).\label{antiT}
\eea
advocated in \cite{Strominger:2013jfa} and derived in the context of graviton scattering in \cite{He:2014laa}, but the parity-even BMS supertranslations $\ell =2,4,\dots$ do not obey the antipodal matching conditions \eqref{antiT}.

\section{Vacua with sources}
\label{vws}

It has been proposed that the asymptotic symmetry group of flat spacetime should be enhanced to include superrotations \cite{Barnich:2009se,Barnich:2011ct} (see also \cite{Banks:2003vp}). In turn, consistency of the algebra would then impose that the symmetry group should as well contain supertranslations generated by singular functions over the sphere \cite{Barnich:2009se}. If that proposal is correct, there should exist a phase space obtained by the exponentiation of these infinitesimal diffeomorphisms which admit all usually accepted physical properties such as positivity of the energy and existence of a variational principle for the vacua. 

In the following, we shall give the result of this exponentiation and provide with the resulting phase space. However, we will show that these two fundamental physical properties are violated: energy is unbounded from below and there is no variational principle. We will conclude that the superrotations should not be considered as a part of the symmetries of Einstein gravity. In turn, this implies that there is no physical Goldstone boson associated with this symmetry. Instead, we will be led to understand the ``would be Goldstone boson''  as a source which is fixed to zero as a result of boundary conditions. 

\subsection{Metric}

After some lengthly algebra, one can exponentiate the superrotation generators at the same time as the supertranslation generators in order to find a metric labelled by two functions: the function $C(z,\bar z)$ generated by the finite supertranslation diffeomorphism and the function $G(z)$ and its complex conjugate $\bar G(\bar z)$ generated by the finite superrotation diffeomorphism. By construction, $G(z)$ is a meromorphic function the sphere and $C(z,\bar z)$ is an arbitrary function on the sphere with possible poles. 

The resulting metric which we call the vacua with sources can be written as 
\bea
ds^2 =\left[\begin{matrix}
-1-\frac{\p_u U}{\sqrt{r^2+U}}&-\frac{r}{\sqrt{r^2+U}}&\frac{1}{2}D^BC_{AB}-D_A\sqrt{r^2+U}&\\-\frac{r}{\sqrt{r^2+U}}&0&0&\\\frac{1}{2}D^BC_{AB}-D_A\sqrt{r^2+U} &0&(r^2+2U)\gamma_{AB}+C_{AB}\sqrt{r^2+U}&\\
\end{matrix}\right] \label{vacuaRot}
\eea
where $U=\frac{1}{8}C_{AB}C^{AB}$ and
\bea
C_{AB}= C^T_{AB} +2 (u + C) T_{AB}.\label{defCtot}
\eea
Here
\bea
C^T_{AB}=-(2D_AD_B-\gamma_{AB} D^2)C
\eea
is the same tensor on the sphere as the one defined for the vacua. The new tensor $T_{AB}$ is traceless and divergence-free. Its holomorphic components are defined as
\bea
T_{zz} = -\frac{1}{2}S[G(z),z]
\eea
where $S[G(z);z]$ is the Schwarzian derivative of $G$
\bea
S[G(z);z] = \frac{\p_z^3G}{\p_zG }-\frac{3 (\p^2_zG)^{2}}{2 (\p_zG)^{2}}. 
\eea
Its antiholomorphic components are defined analogously and $T_{z \bar z} = 0$.

The occurance of a Schwarzian derivative is familiar from $3$ dimensional examples in $AdS_3$ which also possess local conformal invariance (see e.g. \cite{Balog:1997zz}). 
It is natural to define the free real boson on the stereographic plane $\Psi$ as $\Psi = \psi(z) + \bar \psi(\bar z)$ where 
\bea
\p_z G = e^\psi,\qquad \p_{\bar z} \bar G = e^{\bar \psi}.
\eea
Then the traceless and divergence-free tensor on the sphere $T_{AB}$  is recognized as the stress-tensor of the free boson $\Psi$, 
\bea
T_{zz} = \frac{1}{4} (\p_z \Psi)^{2}-\frac{1}{2}\p_z \p_z \Psi ,\qquad T_{\bar z \bar z} = \frac{1}{4}(\p_{\bar z} \Psi)^{2}-\frac{1}{2}\p_{\bar z} \p_{\bar z} \bar\Psi . 
\eea
As in the toy model with $AdS_3$ asymptotics, one can generalize the solution by adding zero modes $\Lambda$ and $\bar \Lambda$, which are not obtained by coordinate transformations. We define the  Liouville stress-tensor by completing the stress-tensor of the free boson as
\bea
T_{zz} = \frac{1}{4} (\p_z \psi)^{2}-\frac{1}{2}\p_z \p_z \psi +\frac{\Lambda}{4} e^{2 \psi} ,\qquad T_{\bar z \bar z} = \frac{1}{4}(\p_{\bar z} \bar \psi)^{2}-\frac{1}{2}\p_{\bar z} \p_{\bar z} \bar\psi+\frac{\bar \Lambda}{4}  e^{2 \bar \psi}. 
\eea
We still define the real boson $\Psi$ as $\Psi = \psi(z) + \bar \psi(\bar z)$. One can check that the solution is still Riemann flat everywhere in the coordinate patch.

\subsection{Properties}

Let us first give some general properties of this metric. The metric is locally Riemann flat. In the case $\Lambda = \bar \Lambda = 0$ it is obvious since it is obtained from a finite diffeomorphism applied on Minkowski spacetime. There is again a defect in the geometry signaled by the existence of conserved charges, see Section \ref{charges}. 

Under the action of an infinitesimal supertranslation, the fields transform as 
\bea
\delta_T C = T(z,\bar z),\qquad \delta_T T_{AB} = 0. 
\eea
The infinitesimal diffeomorphism which generates superrotation is given by 
\bea
\chi_R = \frac{1}{2}uD_{A}R^A\partial_u+(-\frac{1}{2}(r+u)D_AR^A+\mathcal{O}(\frac{  1}{r})\partial_r+(R^A-\frac{u}{2r}D^AD_BR^B+\mathcal{O}(\frac{1}{r^2}))\partial_A. \label{chiR1}
\eea
Here $R^A=(R(z),\bar R(\bar z))$ are the conformal Killing vectors on the sphere. This vector field is in general singular on the sphere except for the global Killing vectors generated by $R=1,z,z^2$. 
Under the action of an infinitesimal superrotation,  the fields transform as 
\bea
\delta_R C &=&R \p_z C - \frac{1}{2} C \p_z R + C \p_z \phi R + c.c., \\
\delta_R T_{zz}  &=& R \p_z T_{zz}+2 \p_z R T_{zz} -\frac{1}{2} \p_z^3 R,\label{dT}\\
\delta_R T_{\bar z \bar z}  &=& \bar R \p_{\bar z} T_{\bar z \bar z}+2 \p_{\bar z} \bar R T_{\bar z \bar z} -\frac{1}{2} \p_{\bar z}^3 \bar R.
\eea
Here we recall that $\phi$ is defined from the conformal factor on the sphere as $e^{-2\phi} = \gamma_{z \bar z}$. 
The transformation law for $T_{AB}$ is exactly the one of the stress-tensor of a $2d$ Euclidean CFT. The holomorphic and anti-holomorphic bosons $\psi,\, \bar \psi$ simply transform as 
\bea
\delta_R \psi = R \p_z\psi + \p_z R,\qquad \delta_R \bar \psi = \bar R \p_{\bar z}\bar \psi + \p_{\bar z} \bar R.
\eea

\subsubsection{Energy unbounded from below}

The metric is written in BMS gauge at future null infinity (see appendix \ref{BMSg} for the definition of BMS gauge).
The canonical mass associated with $\p_t$ can be computed using the Barnich-Brandt method \cite{Barnich:2001jy}. One finds that it is not integrable and that the infinitesimal mass variation $\slash\hspace{-6pt}\delta M$ can be written as the sum of the variation of the Bondi mass $\delta M_B$ and a $u$-independent non-integrable piece. So even though the mass is not integrable, the time derivative of the mass is integrable and is given by the time derivative of the Bondi mass. One can easily read off the Bondi mass at future null infinity which is given by 
\bea
M_B &=& -\frac{1}{2}\p_u U = -\frac{1}{8}N_{AB}C^{AB} \\
&=&- \frac{1}{4}T_{AB}C_T^{AB} - \frac{1}{2}(u+C) T^{AB}T_{AB}.
\eea
The Bondi mass decreases with retarded time $u$ as it should,
\bea
\p_u M_B &=& - \frac{1}{2}T^{AB}T_{AB}.
\eea
However, since the vacua have zero energy, the presence of $T_{AB}$ is unphysical. It drives the energy towards minus infinity. We should therefore interpret the presence of $T_{AB}$ as a source which ought to be set to zero for physical field configurations\footnote{Also note that the presence of constant news $N_{AB} \equiv \p_u C_{AB}= 2 T_{AB}$ leads to a violation of the Christodoulou-Klainerman asymptotic flatness conditions which instead require that the news decays as  $N_{AB} \simeq |u|^{-3/2}$ for large $u \rightarrow \pm \infty$ \cite{Christodoulou:1993uv}. }. The set of metrics \eqref{vacuaRot} is certainly a solution space but it should not be interpreted as a set of solutions obeying physical boundary conditions. Given this observation, we are led to impose the set of Dirichlet boundary conditions
\bea
T_{AB} = 0.\label{DBC}
\eea
The role of these boundary conditions will become clearer in the next section.

\subsubsection{Symplectic flux originating from sources}
\label{varp}

Let us now study the existence of a variational principle for the solution space \eqref{vacuaRot} which will lead to the definition of symplectic flux at $\mathfrak I^+$. It will clarify the role of the Dirichlet boundary conditions and the interpretation of $T_{AB}$ as a source.

The action of Einstein gravity is
\bea
S = \frac{1}{16 \pi G} \int_{\mathcal M} d^4 x \sqrt{-g}R + \int_{\mathfrak{I}^+} B[g] + \dots
\eea
It admits a boundary term close to future null infinity. It is defined as usual on a finite and large radial cutoff hypersurface $r = \lambda $ (in BMS coordinates) with $\lambda \rightarrow \infty$. It might depend  upon the intrinsic and extrinsic fields of the induced metric on the hypersurface. In order to keep notations short, we just express the codimension 1 boundary form $B[g]$ as a function of the bulk metric.
It might contain the Gibbons-Hawking term or other terms. There are also additional boundary terms at spacelike and past null infinity indicated by the dots but we shall not be concerned about them here.  

Note that the surface $r = \lambda $ admits a normal which is slighly spacelike, so the surface is slightly timelike. It has therefore the same signature as a cutoff surface close to the boundary of AdS, for which the variational problem is very well understood \cite{Balasubramanian:1999re}.

The variation of the action is given by a bulk term proportional to the equations of motion plus a boundary term,
\bea
\delta S &=& - \frac{1}{16 \pi G}\int_{\mathcal M} d^4 x  \sqrt{-g}G^{\mu\nu}\delta g_{\mu\nu} + \int_{\mathfrak{I}^+} du dz d\bar z\left( \Theta^r[\delta g ; g] + \delta B[g] \right) + \dots 
\eea
Here, the presymplectic potential form $ {\boldsymbol \Theta}[ \delta g ; g] = \Theta^\mu [ \delta g ; g](d^3 x)_\mu$ is the codimension 1 form defined from varying the Einstein-Hilbert action and obtained after performing integration by parts,
\bea
\delta (\sqrt{-g} R) d^4x= -\sqrt{-g}G^{\mu\nu}\delta g_{\mu\nu} d^4x +\p_\mu \Theta^\mu[ \delta g ; g].\label{Theta}
\eea

Let us now specialize to our setting with $g$ an element of the solution space \eqref{vacuaRot} and $\delta g$ tangent to the solution space \eqref{vacuaRot}. We obtain 
\bea
\Theta^r[\delta g ; g] &=&\frac{1}{16 \pi G} \sqrt{\gamma} (-C_{AB} \delta T^{AB}-\frac{1}{2}D_AD_B\delta C^{AB})+ O(r^{-1}).
\eea

The existence of a variational principle amounts to the existence of a suitable boundary term $B[g] $ such that the variation of the action is zero. The integrability condition for solving this problem is exactly the requirement that the following boundary symplectic structure be zero,
\bea\label{symstr}
\Omega_{\mathfrak I^+}[\delta C,\delta \psi ; \delta C,\delta \psi ] \equiv -\frac{1}{4 G}  \int_{\mathfrak I^+} du d^2\Omega \, \delta C_{AB} \wedge \delta T^{AB}.
\eea

We require that the vacua  admit no symplectic structure at $\mathfrak I^+$ since it would be incompatible with their stability. This condition is exactly equivalent to the existence of a variational principle for the vacua. One can check that the boundary symplectic structure at $\mathfrak I^+$ is non-vanishing for generic variations, even around the vacua $T_{AB} = 0$\footnote{As an explicit example, let us set $\delta C$ to be the $m=l=2$ spherical harmonic function and $\delta T_{\bar z \bar z}$ a constant (generated by $\bar R = \bar z^3$) . Then $\delta C_{zz}(\gamma^{z\bar{z}})^2 \delta T_{\bar z \bar z}$ is a constant so the integral on the sphere is nonvanishing.}. The absence of symplectic flux at $\mathfrak I^+$ or the existence of a variational principle therefore requires a boundary condition. Requiring energy bounded from below singles out the Dirichlet boundary condition \eqref{DBC}. 

Let us now further interpret \eqref{symstr}. We recognize that $T^{AB}$ and $C_{AB}$ are canonically conjugated variables. 
 $C_{AB}$ can take a vacuum expectation value as a result of spontaneous BMS supertranslation symmetry breaking. Its conjugated variable $T^{AB}$ is its conjugated source which is fixed as a result of Dirichlet boundary conditions. We therefore found a dictionary which is strikingly similar to the AdS/CFT dictionary at the boundary of AdS where conjugated variables come in pairs of a source and a vacuum expectation value. 

Since superrotations turn on $T_{AB}$, these transformations do not lead to metric perturbations tangent to the physical phase space.  Nevertheless, superrotations still have a role to play as we will see in the following as external symplectic symmetries. As in AdS/CFT, one is allowed to turn on sources infinitesimally in order to probe the physical state, as we will discuss.

\section{Symplectic symmetries of the vacua}
\label{symplectic}

The concept of symplectic symmetry in gravitational and gauge theories has been put forward in \cite{Compere:2015knw,Compere:2015bca} based on the earlier observations in \cite{Barnich:2010eb,Compere:2014cna}. 
A symplectic symmetry is defined as a gauge parameter (an infinitesimal diffeomorphism $\chi^\mu$ in the case of gravity) which is not an exact symmetry (a Killing symmetry in the case of gravity) such that it exists a presymplectic form which is zero on-shell when contracted with the corresponding gauge transformation
\bea
{\boldsymbol \omega}[ \mathcal L_\chi g, \delta g; g] \approx 0. \label{symp}
\eea
Here $g$ labels an arbitary point in the phase space and $\delta g$ any metric tangent to the phase space.
A symplectic symmetry is canonically associated with a conserved charge which only depends upon the homology of the sphere of integration $S$ (it preserves its value upon smoothly deforming the sphere).
Indeed, one has the following fundamental theorem which could be dubbed ``the generalized Noether theorem for gauge theories'' \cite{Iyer:1994ys,Barnich:1995ap,Barnich:2000zw,Barnich:2001jy,Barnich:2007bf}:
\bea
{\boldsymbol \omega}[ \mathcal L_\chi g , \delta g ; g]  = d {\boldsymbol k}_\chi [ \delta g ; g] +(EOM) + \delta (EOM). \label{GN}
\eea
Here equality holds up to terms vanishing on-shell ($g$ obeying Einstein's equations) and linearized on-shell ($\delta g$ obeying the linearized Einstein's equations). It will be important for the following that the surface charge form ${\boldsymbol k}_\chi [ \delta g ; g] $ is defined uniquely from the presymplectic form up to an exact term $d {\boldsymbol l}_\chi[\delta g ; g]$ where ${\boldsymbol l}_\chi$ is a codimension 3 form. The conserved surface charge is then defined as
\bea
\mathcal Q_\chi = \int_S \int_{\bar g}^g  {\boldsymbol k}_\chi [ \delta g' ; g'] . 
\eea
It is independent of the path chosen in phase space if the so-called integrability conditions are obeyed. A symplectic symmetry is then called \emph{trivial} if its conserved charge is always zero. It is called a \emph{non-trivial} symplectic symmetry otherwise. The two independent definitions for the presymplectic form in Einstein gravity and their associated conserved surface charges are recalled in appendix \ref{App:Symp}.


In the following we will present a prescription for fixing the ambiguities in the Lee-Wald presymplectic form in order to obtain a presymplectic form which exactly vanishes for any vacuum \eqref{vacua} and any perturbation $\delta g$ generated by both supertranslations \emph{and} superrotations. Supertranslations will therefore generate symplectic symmetries of the vacua. Superrotations are not tangent to the phase space of vacua as we discussed. We will show nevertheless that they can be defined as \emph{external} symplectic symmetries, which are also associated with conserved charges in the physical phase space.

\subsection{Derivation of the on-shell vanishing presymplectic form}

Our main interest is the space of vacua \eqref{vacua} and the perturbations $\delta_T g$, $\delta_R g$ generated by supertranslations and superrotations. The natural setting to study is therefore the solution space of vacua with sources $g(C,\psi)$ \eqref{vacuaRot} since supertranslations act tangentially on that solution space. The solution space with sources is labelled by the real function $C(z,\bar z)$ (with poles) and the meromorphic function $\psi(z)$ on the sphere. 

Let us first compute the two standard definitions of the presymplectic form given in Appendix \ref{App:Symp} in our setting. We obtain that neither the Lee-Wald neither the invariant presymplectic forms are vanishing. 
There is however a well-known ambiguity in the definition of the Lee-Wald presymplectic form that we could exploit as recalled in Appendix \ref{App:Symp}. We therefore aim to find a codimension 2 form $ {\boldsymbol \Theta}_{ct}[ \delta g ; g] $ such that the total presymplectic form 
\bea
{\boldsymbol \omega} [\delta_1 g , \delta_2 g ; g] = {\boldsymbol \omega}_{LW} [\delta_1 g , \delta_2 g ; g] -d \left(  \delta_1 {\boldsymbol \Theta}_{ct}[ \delta_2 g ;g]    -  \delta_2  {\boldsymbol \Theta}_{ct}  [\delta_1 g ; g]  \right),\label{totomega}
\eea
is vanishing for any vacua $g(C)$ \eqref{vacua} and any perturbations  $\delta_T g$, $\delta_R g$. 

Let us start by deriving the asymptotic behavior of the presymplectic potential form $\Theta^\mu[\delta g ; g(C,\psi)]$. It obeys 
\bea
\Theta^u[\delta g ; g(C,\psi)] &=& 0,\nn\\
\Theta^r[\delta g ; g(C,\psi)] &=& \Theta_{(0)}^r[\delta g ; g(C,\psi)] + O(r^{-1}),\label{aT}\\
  \Theta^A[\delta g ; g(C,\psi)] &=& O(r^{-2}). \nn
\eea
The finite radial component was already found in Section \ref{varp} and reads as
\bea
\Theta_{(0)}^r[\delta g ; g(C,\psi)]  = \frac{1}{16 \pi G} \sqrt{\gamma} (-C_{AB} \delta T^{AB}-\frac{1}{2}D_AD_B\delta C^{AB}). 
\eea
Let us recall that one defines the presymplectic potential $ {\boldsymbol \Theta}[ \delta g ; g] $ from varying Einstein's equations, $\delta (\sqrt{-g} R) = -\sqrt{-g}G^{\mu\nu}\delta g_{\mu\nu}+\p_\mu \Theta^\mu[ \delta g ; g] $. By Einstein's equations and linearized Einstein's equations, the left-hand side is zero on-shell and therefore ${\boldsymbol \Theta}$ is conserved, $\p_\mu \Theta^\mu= 0$. One can therefore write $\Theta^r$ exactly as
\bea
\Theta^r =\Theta_{(0)}^r[\delta g ; g(C,\psi)]  - \int dr  \p_A \Theta^A[\delta g ; g(C,\psi)] .
\eea
After varying the fields in \eqref{aT}, this leads to the following exact expression for the Lee-Wald presymplectic form
\bea
\omega_{LW}^u = 0,\qquad
\omega_{LW}^r =\omega_{(0)}^{r} +  \p_A \delta \left( - \int dr \Theta^A \right),\qquad
\omega_{LW}^A = \p_r \delta \left(  \int dr \Theta^A \right).
\eea
We rewrote $\omega_{LW}^A$ in a suggestive form and defined
\bea
\omega_{(0)}^{r} [\delta_1 g,\delta_2 g ; g(C,\psi)]  \equiv -\frac{1}{16 \pi G} \, \sqrt{\gamma}\left( \delta_1 C_{AB} \, \delta_2 T^{AB} -(1 \leftrightarrow 2)  \right). 
\eea
Therefore, we were able to write the Lee-Wald presymplectic form as an exact differential of a codimension 2 form, up to the remaining constant piece $\omega_{(0)}^{r}$,
\bea
\boldsymbol \omega_{LW} = \boldsymbol \omega_{(0)} + d (\delta_1    {\boldsymbol \Theta}_{ct}(\delta_2 g ; g) - \delta_2    {\boldsymbol \Theta}_{ct}(\delta_1 g ; g) )
\eea
where
\bea
\Theta^{u A}_{ct} = \Theta^{AB}_{ct}  = 0, \qquad \Theta^{rA}_{ct} = -\Theta^{Ar}_{ct}  = \int dr \; \Theta^A.
\eea
Here, we defined for convenience the other components of $ \boldsymbol \omega_{(0)}$ as $\omega_{(0)}^u=0$, $\omega_{(0)}^A=0$. 

Now, around a generic element of the vacua with sources, the presymplectic form $ \boldsymbol \omega_{(0)}$  cannot be written as an exact differential of an extra field variation, i.e. of the form $d\,\delta (\cdot)$. The offending term is proportional to $\delta_1 C \delta_2 (T_{AB}T^{AB})- (1 \leftrightarrow 2)$ after using \eqref{defCtot}\footnote{One might define $\Theta_{ct}^{ru} \propto u C \delta (T_{AB}T^{AB})$ after integrating over $u$ but such terms correspond to the addition of a compensating energy flux originating from $\mathfrak I^+$ going inwards, which is not admissible. One could also write $\delta g$ as a linear combination of variations caused by supertranslations and superrotations and use the generalized Noether theorem \eqref{GN} to write $ \boldsymbol \omega_{(0)}$ as a boundary term. However, it will not be of the form $d\,\delta (\cdot)$ since the charges are not integrable in general.}. However, we now observe that around the physical space of vacua where $T_{AB}$ (but not its variation) is constrained by the Dirichlet boundary condition \eqref{DBC} the offending terms are absent and the presymplectic form $ \boldsymbol \omega_{(0)}$ can be written as a boundary term which is an exact field variation! It reads as 
\bea
\boldsymbol \omega_{(0)} = \frac{1}{2G} du d^2\Omega \, D_A \left(  D_B \delta_1 C\, \delta_2 T^{AB} -(1 \leftrightarrow 2)  \right) \label{sym2}
\eea
after using the properties that $T^{AB}$ is traceless and divergence-free. Therefore, for any perturbation around the vacua without sources we have
\bea
\boldsymbol \omega_{LW} = \boldsymbol d (\delta_1    {\boldsymbol \Theta}_{ct}(\delta_2 g ; g) - \delta_2    {\boldsymbol \Theta}_{ct}(\delta_1 g ; g) )
\eea
where $\Theta^{\mu \nu}_{ct}  = \Theta^{[\mu \nu]}_{ct} $ and
\bea
\Theta^{u A}_{ct} &=& \Theta^{AB}_{ct}  = 0, \\
\Theta^{rA}_{ct} &=&  \frac{1}{8 \pi G} \sqrt{\gamma} \,  D_B  C \delta T^{AB}   + \int dr \; \Theta^A.
\eea
With this definition, the total presymplectic form \eqref{totomega} is identically zero for any perturbations $\delta_T g$, $\delta_R g$ around the vacua \eqref{vacua}.

Supertranslations are therefore symplectic symmetries and lead to conserved charges for any codimension 2 surface enclosing a singularity such as a defect. The superrotations lead to perturbations which are not tangent to the phase space. Yet, we have the property
\bea
{\boldsymbol \omega}[\delta_R g, \delta_{T} g; g(C) ] = 0
\eea
which leads to the existence of closed surface charge forms $ {\boldsymbol k}_{\chi_R}$ and consequently conserved charges around the vacua thanks to the generalized Noether theorem \eqref{GN}. We will call the superrotations \emph{external symplectic symmetries} to distinguish them from the supertranslations which we could call \emph{internal symplectic symmetries}. With that definition both external and internal symplectic symmetries lead to conserved charges but only the internal ones are tangent to the physical phase space.

Usually transformations which do not preserve the phase space are not associated with conserved charges. Here however, something very peculiar happened.
The symplectic structure at $\mathfrak I^+$ \eqref{symstr} is non-zero, as we discussed, so superrotations are not canonical:  they do not generate perturbations tangent to the phase space. However, the presymplectic form is actually a boundary term for variations around the physical configurations $T_{AB}=0$. Usually such a boundary term would be zero but here it is not, because of the poles of the meromorphic and anti-meromorphic functions $\delta T_{zz}$ and $\delta T_{\bar z \bar z}$. (We will work out examples of integrals with poles in the following). 

\subsection{Conserved BMS supertranslations and superrotations}
\label{charges}

Let us now turn our attention to the conserved charges of the vacua associated with supertranslations and superrotations. The on-shell vanishing presymplectic form leads via the generalized Noether theorem to a  closed surface charge form $ {\boldsymbol k}_{\chi}$ which can be integrated on any codimension 2 surface $S$. More precisely, the surface charge form is defined as 
\bea
{\boldsymbol k}_{\chi}  [ \delta g ; g(C)]  = {\boldsymbol k}^{IW}_{\chi} [ \delta g ; g(C)] +{\boldsymbol k}^{ct}_{\chi} [ \delta g ; g(C)]  +d {\boldsymbol l}_\chi [ \delta g ; g(C)] .\label{defk1}
\eea
The Iyer-Wald surface charge ${\boldsymbol k}^{IW}_{\chi}$ originates from the Lee-Wald presymplectic form as explained in Appendix \ref{App:Symp}. The supplementary term ${\boldsymbol k}^{ct}_{\chi}$ originates from the additional boundary term in the presymplectic form \eqref{totomega} as reviewed in Appendix \ref{App:Symp}. Since only the $rA$ components of $\boldsymbol \Theta_{ct}$ are non-vanishing, the same is true for ${\boldsymbol k}^{ct}_{\chi}$. When the surface $S$ is the sphere at constant $u,r$, such a term does not contribute. It would contribute for a more generic codimension 2 surface $S$ of integration.
The final term is the exact differential of a codimension 3 \emph{line charge form} which is unfixed by the theory. Usually, one completely ignores it. Here however, one cannot due to singularities at the poles of the supertranslation function $R(z)$ which can lead to non-vanishing integrals $\oint_S d {\boldsymbol l}_\chi$. We already saw an example of such non-vanishing integrals of an exact differential over the sphere: the symplectic structure at $\mathfrak I^+$ around the vacua without sources, see \eqref{symstr} and \eqref{sym2}.

Let us consider a surface $S$ surrounding the locus $r=0$. The infinitesimal conserved charge associated to the generator $\chi$ is defined as 
\bea
Q^S_\chi = \oint_S \int_0^{C(z,\bar z)} {\boldsymbol k}_\chi [dg ; g(C)] .
\eea
The integration is performed first  in the phase space between the global Poincar\'e vacuum and the vacuum of interest characterized by the Goldstone boson $C(z,\bar z)$. The one-form in phase space $dg$ is defined as $dg_{\mu\nu} = \frac{\p g_{\mu\nu}}{\p C} dC$. The resulting surface form is then integrated in spacetime over the surface $S$.

It is straightforward to compute the conserved charge associated with supertranslations on the sphere around future null infinity. One has $\oint_S d {\boldsymbol l}_\chi = 0$ by Stokes' theorem applied on regular functions and we find the vanishing result
\bea
\mathcal Q^S_{\chi_T}  = 0.
\eea
The surface $S$ can also be smoothly deformed at any radius and the charge still be exactly 0. All vacua therefore admit zero energy and momentum and zero BMS supertranslation charge. These symplectic symmetries are therefore trivial. The supertranslation symmetry breaking leads to the existence of a non-trivial Goldstone boson but this boson does not contribute to the supertranslation charges. This result agrees with a similar computation made in \cite{Flanagan:2015pxa} in the context of stationary spacetimes. This result is also in line with the zero supertranslation charges found on the horizon of a stationary black hole \cite{Hawking:2016msc}.

Let us now consider the superrotations. The generator of superrotation was given in \eqref{chiR1}. It depends upon a meromorphic function $R(z)$. The poles of the function $R(z)$ have an important consequence: the integral of an exact differential $d {\boldsymbol l}_{\chi_R}$ (linear in $\chi_R$) over the sphere might be non-vanishing. 

The known theory of conserved charges however gives no information on how to fix these codimension 3 line charge forms $ {\boldsymbol l}_{\chi_R}$. We will therefore need a physically motivated prescription to define the surface charge \eqref{defk1} associated with superrotations. As a part of our prescription, we require that the superrotation charge is exactly zero for global Minkowski spacetime. Since $C_{AB}= 0$ for the global vacuum, we require that the superrotation charge depends upon $C$ only through $C_{AB}$.  We will see however that this requirement is not enough to prove uniqueness of the definition of the superrotation charge.

Let us start by evaluating the Iyer-Wald charge on the sphere at infinity as defined in Appendix \ref{App:Symp}. We find that the charge diverges linearly in $r$ but that divergent term can be absorbed by an exact differential of a line charge form. The charge is then finite but linear in $u$. Again, this $u$ dependent term can be absorbed by a line charge form. We then note that the finite time independent result is integrable.  
The superrotation charge is finally obtained as 
\bea
\mathcal Q^S_{\chi_R}=-\frac{1}{4G}\oint_S d^2\Omega R^A(\frac{1}{8}D_A(C_{EF}C^{EF})+\frac{1}{2}C_{AB}D_{E}C^{EB}).\label{QR}
\eea
This expression exactly matches with the one derived by Barnich-Troessaert \cite{Barnich:2011mi}. Indeed, in the absence of news, the formula for the superrotation charge proposed in \cite{Barnich:2011mi} is integrable and reads as
\bea
\mathcal Q^S_{\chi_R} = \frac{1}{4 G}\oint_S
   d^2 \Omega \, R^A \Big[ \big(2 N_A +
\frac{1}{16}\partial_A (C^{CB} C_{CB})\big) \Big]\,,\label{GT}
\eea
where $d^2 \Omega$ is the unit measure on the round sphere and $N_A$ is defined in Appendix \ref{BMSg}. After substituting the explicit expressions for the vacua, the charge \eqref{GT} exactly reproduces \eqref{QR}.\footnote{Consistently with that matching, we also checked that the term $\boldsymbol E[\delta_R g ; \delta_T g ; g(C)]$ which differs between the Barnich-Brandt charge and the Iyer-Wald charge is zero on the sphere at infinity. (See appendix \ref{App:Symp} for definitions.)}

It turns out that the charge \eqref{QR} enjoys the required physical property: it is identically zero for the Minkowski global vacuum since it is a function of $C_{AB}$ which vanishes on the global vacuum. However, one could as well perform integrations by parts of the remaining derivative in the expression \eqref{GT} while keeping the structure of $C_{AB}$ intact and the resulting charge will still obey the physical requirement. In the lack of a more precise prescription, we adopt the definition of the superrotation charge to be \eqref{QR}.

It is important to check that any regular function $C(z,\bar z)$ leads to finite superrotation charges. We checked that setting $C(z,\bar z)$ to be an arbitrary combination of the lowest non-trivial harmonics $l=2$, $m=-2,\dots 2$ and any $R(z) = z^k$, $k \in \mathbb Z$ the superrotation charges are finite. Since nothing special happens for higher harmonics or at other locations than $z=0$, we then argue that the charges are always finite. More precise mathematical results are known for the integrals of functions with poles \cite{Dickenstein:1999gb}. A crucial ingredient is the residue theorem: any meromorphic function on the sphere is such that the sum of its residues vanishes.

Since the superrotation charges might be non-vanishing, there ought to be a defect at $r=0$ at constant $u$ or beyond that locus that prevents the charge to be zero after shrinking the sphere $S$ to a point. This proves the existence of a defect in the bulk spacetime, as announced earlier. It also proves that the superrotation external symplectic symmetries are non-trivial.

An important property of the superrotation charge is that the rotation and Lorentz charges are all vanishing for any $C(z,\bar z)$. In order to prove this statement, one first needs to rewrite the charge formula \eqref{QRO} as
\bea
\mathcal Q^S_{\chi_R}&=&-\frac{1}{8G}  \oint_S d^2\Omega(D_AD_B+\gamma_{AB})D_ER^E(D^ACD^BC).\label{QRO}
\eea
The proof of the equivalence of \eqref{QR} and \eqref{QRO} for the Lorentz group is given in Appendix \ref{proof1} and uses the fact that we can freely drop boundary terms when considering globally defined vectors. Explicitly, in the $z,\bar z$ coordinate system,
\be
\mathcal Q^S_{\chi_R}=-\frac{1}{8G}\oint_S d^2\Omega (\partial_z^3R(z)) (D^zC)^2+c.c.
\ee
Since $R$ is a complex linear combination of $1,z,z^2$, we obtain that all Poincar\'e charges are identically zero. This proves that the vacua are all Poincar\'e invariant.

Let us now illustrate that the superrotation charges cannot be defined as \eqref{QRO} in the case of singular functions $R(z)$. The global Poincar\'e vacuum is obtained upon picking $C= \sin \theta \sin \phi = i\frac{\bar z-z}{1+z \bar z}$ which leads to $C_{AB} = 0$. However, the charge \eqref{QRO} is non-vanishing which violates our physical criteria! More precisely the superrotation charges associated with $R=z^{-1}$ and $R=z^3$ are respectively $-\frac{3}{8G}$ and $\frac{3}{8G}$. Since the difference between the integrands in \eqref{QRO} and \eqref{QR} is an exact differential, as proven in Appendix \ref{proof1}, we found an example of non-vanishing integral of an exact differential. 

In the AdS/CFT correspondence, sources are used as probes to compute correlators of physical observables \cite{Witten:1998qj}. The situation is similar in flat space holography: the superrotations act as sources which define conserved charges that probe the physical observable: the expectation value of the Goldstone boson $C$.

\section{Multi-centered vacua}
\label{multi}

The vacua that we constructed admit a defect which is signaled by superrotation charges. Since superrotations are symplectic symmetries, one can define the superrotation charges as an integral over a small sphere which surrounds the defect. 

Let us now describe an algorithm to define a vacuum with multiple defects. One starts with the single defect vacuum in $(u,r,z,\bar z)$ coordinates \eqref{vacua} and one defines Cartesian coordinates $x^i$ from the standard change of coordinates from spherical to Cartesian coordinates. One then acts with the diffeomorphism $x^i \rightarrow x^i + a^i$. This diffeomorphism is not associated with a Killing vector. Indeed, translations admit components $\chi^\mu$ depending upon the Goldstone boson $C(z,\bar z)$ since they are obtained from the vector transformation law of the standard translations with a diffeomorphism which depends upon $C(z,\bar z)$. This latter diffeomorphism is precisely the one which generates the single center vacua from the global Poincar\'e vacuum. The diffeomorphism $x^i \rightarrow x^i + a^i$ asymptotes to a translation at future and past null infinity and therefore it is an admissible transformation (it preserves the standard asymptotic flatness boundary conditions). Its effect is to move the locus  $r = 0$ to $x^i = a^i$. One can then rewrite the metric in new spherical coordinates whose origin $r=0$ is off-centered with respect to the previous locus which contains the defect.

As a second step, one acts on the resulting vacuum with a defect at/beyond $x^i = a^i$ with the \emph{same} diffeomorphism as the one used to define the vacuum with defect at/beyond $r=0$ from the global Minkowski vacuum.  The effect of this diffeomorphism is to generate a new defect at/beyond $r=0$ in addition to the already existing one at/beyond $x^i = a^i$. One can then iterate the procedure to obtain a multi-centered vacua. Each defect will be characterized by its own superrotation charges. The superrotation charge associated with a surface enclosing all defects will be the superrotation charge computed at future null infinity. The number of Killing symmetries of the vacua is still 10 since they are only transformed by the diffeomorphisms. Since the multi-centered vacua all admit the same asymptotics they admit zero Poincar\'e charges since one can compute them at infinity as in the single defect case. 

The fact that the superrotation charges can be computed individually is a feature of Killing and symplectic symmetries. A familiar example in the case of Killing symmetry is the following: the electric charges of isolated electrons can be individually computed by enclosing spheres. This feature can be generalized to arbitrary Killing symmetries \cite{Barnich:2003xg} and symplectic symmetries \cite{Compere:2015knw}.

In terms of the Goldstone boson $C$ of spontaneously broken BMS supertranslation invariance, the multi-centered vacua can be interpreted as resulting from a Goldstone boson depending upon the three spatial directions  $C=C(x^i)$. Asymptotically, one still has $C = C_{(0)}(z,\bar z)+O(r^{-1})$ and the function $C_{(0)}$ is characterized by the superrotation charges defined at $\mathfrak I^+$ and $\mathfrak I^-$ which enclose all the defects. 

Let us finally note that in three dimensional analogue models, the multi-centered defects due to conical deficit angles were built and analysed in \cite{Deser:1983tn}. We expect that one can similarly build defects to admit charges under the three-dimensional BMS group \cite{Ashtekar:1996cd,Barnich:2006av}.

\section{Conclusion and Discussion}
\label{ccl}

In mathematical terms, we built the orbit of Minkowski spacetime under the (original) BMS group. The resulting set of metrics are Poincar\'e vacua in the sense that they are Poincar\'e invariant and admit vanishing Poincar\'e charges. They were built from a finite bulk diffeomorphism which exponentiates the infinitesimal supertranslation BMS symmetries at future null infinity. We showed that they admit finite and non-vanishing superrotation charges which can also be extended in the bulk (i.e. they are symplectic symmetries). This then signals the existence of individual bulk defects which source the charges. We also constructed multi-centered vacua which admit individually conserved superrotation charges. We interpreted the supertranslation field which sources the superrotation charges as a finite Goldstone boson associated with broken BMS supertranslation invariance.

We also considered the possibility of defining a Goldstone boson associated with broken BMS superrotation invariance. More precisely, we constructed the orbit of Minkowski spacetime under finite superrotations, which we referred to as vacua with sources. However, we found several pathologies: the energy would then be unbounded from below, there would be a symplectic flux at future null infinity leading to stability issues, the variational principle would not exist and the Christodoulou-Klainerman boundary conditions would be violated. We concluded that the Einstein action does not admit the extended BMS symmetry with superrotations included as asymptotic symmetry group. Instead, we identified finite superrotations as source generating transformations. Turning on sources infinitesimally as probes defines external symplectic symmetries which can be used to define conserved charges of the Poincar\'e vacua which only depend upon the vacuum expectation value of the Goldstone boson. This is somehow a flat spacetime analogue of the AdS/CFT prescription for computing a vacuum expectation value through sources \cite{Witten:1998qj}.

We will now speculate on how the vacuum metrics that we constructed might be related to physics. We see two different mechanisms for the possible relevance of the vacua: either they are produced in the early universe or they are transformed by the interactions of matter and/or gravitational waves. Let us first discuss the cosmological setting. Well below the cosmological scale set by dark energy, our spacetime can be considered as asymptotically flat. In fact, many patches around us which are sufficiently empty can be considered as asymptotically flat. Such local asymptotically flat patches might have been created originally (let say in an inflationary phase) in a different vacuum than the global vacuum. Indeed, BMS invariance of the action predicts the existence of inequivalent vacua which spontaneously break BMS symmetry. Assuming that our vacuum is the global Minkowski vacuum without further justification could be argued to be equivalent to fine-tuning. 

The existence of a superrotation charge points to an obstruction to shrink the surface of integration in the bulk and suggests a sort of bulk defect. Cosmic censorship prevents the occurance of naked curvature singularities but zero mass cosmic defects are much milder: the Riemann tensor vanishes everywhere, except possibly at the defect.  The nature of these defects is not understood and raises new questions. In $3d$ Einstein gravity with Brown-Henneaux boundary conditions, one can similarly build the Virasoro orbit of the global $AdS_3$ vacuum. Such spacetimes admit conserved finite Virasoro charges in the bulk. There is therefore a similar obstruction at shrinking the circle at infinity to a point due to the presence of these charges. In the case of finite Virasoro descendants of global $AdS_3$, the resulting metric does not describe vacua since the $L_0$ charge changes as a result of the non-abelian algebra. But the presence of bulk defects is similar to the $4d$ asymptotically flat case considered here. If the Poincar\'e vacua that we constructed are physical, we should therefore expect the presence of spacetime defects around us\footnote{Several other possible mechanisms for cosmological spacetime defects have been discussed \cite{Vilenkin:2000jqa,Witten:1985fp,Copeland:2003bj,Hossenfelder:2014hha}.}. The superrotation charge of a sphere enclosing the defect could be measurable in principle by using a ruler, reconstructing the metric and reconstructing the canonical surface charge.

We could build a large degeneracy of classical vacua. How many defects can there be in a given volume? For finite energy particles in a large volume their presence is limited by the existence of black holes. Then the entropy in a finite volume is roughly bounded by its area in Planck units \cite{Bousso:1999xy}. Here, the vacua have zero energy so the problem is more acute and a resolution would require a more detailed knowledge of quantum gravity\footnote{In $AdS_3$, Virasoro descendants of the vacuum can be interpreted as a coherent gas of holographic gravitons or in more mathematical terms as a coherent state in the vacuum Verma module of the dual CFT. Such states can be quantized thanks to the AdS$_3$/CFT$_2$ correspondence. Recent work in that direction includes \cite{Barnich:2014kra,Barnich:2015uva,Oblak:2015sea,Barnich:2015mui,Garbarz:2015lua,Banerjee:2015kcx}.}. One speculation is that the defects become extended objects in the quantum theory such that the sphere surrounding them has a quantized area in terms of Planck cells just sufficient to encode all non-vanishing superrotation quantum bits of information. 

Let us now describe the second possible mechanism where the vacua are relevant. This mechanism is related to interactions of gravitational waves and/or matter. It is well-known that memory effects occur in Einstein gravity, with observable consequences on inertial detectors \cite{Zeldovich:1974aa,Christodoulou:1991cr}. Such memory effects can be attributed to a change of vacuum as recently emphasized in \cite{He:2014laa,Strominger:2014pwa}. While changes in the vacuum lead to physical effects mediated by the Weyl curvature, it is speculative whether or not there is an absolute notion of vacuum. The Poincar\'e vacua that we constructed are candidates for an absolute notion of vacua. They have vanishing Weyl curvature but still admit non-trivial canonical surface charges. Now, it is not clear that stationary finite BMS supertranslations correctly describe the final states of scattering events. One possible issue is that the vacua do not obey the antipodal matching conditions imposed at spatial infinity. However, if the vacua only describe the final state after all transient processes, the extrapolation of the vacua to spatial infinity might be irrelevant. Also, since these antipodal matching conditions were originally obtained for small deviations from Minkowski spacetime, one could argue that they might be modified for spacetimes with finite departure from Minkowski. This remains to be fully understood.

All the analysis can be generalized to the case where the defects lie inside the horizon of a black hole. There is then no naked defect and the vacuum surrounding the black hole is characterized by non-trivial superrotation symplectic symmetries. This might clarify the nature of the black hole states raised in \cite{Hawking:2016msc}. The BMS hair of black holes (i.e. the vacuum expectation value of the Goldstone boson) leads to observables which are the superrotation charges, not the supertranslation charges which identically vanish. The properties of black holes with BMS hair are further addressed in \cite{Compere:2016hzt}.

\section*{Acknowledgments}

The authors would like to thank L. Donnay for her early involvment in the project and G. Barnich, M. Guica, B. Oblak and A. Strominger for their comments on the manuscript and P. Mitra for a useful correspondence. G.C. is Research Associate of the Fonds de la Recherche Scientifique F.R.S.-FNRS (Belgium). G.C. and J.L. both acknowledge the current support of the ERC Starting Grant 335146 ``HoloBHC". This work is also partially supported by FNRS-Belgium (convention IISN 4.4503.15).

\appendix

\section{Computational details}

\subsection{Asymptotic form of the metric}
\label{BMSg}

Any asymptotically flat metric at future null infinity is the sense of Bondi-van der Burg-Metzner-Sachs can be put in the form (see e.g. \cite{Barnich:2011mi})
\begin{equation}
 ds^2=e^{2\beta}\frac{V}{r} du^2-2e^{2\beta}dudr+ g_{AB}(dz^A-U^Adu)(dz^B-U^Bdu),
\end{equation}
where
\bea
g_{AB} &=& r^2 \gamma_{AB} + rC_{AB}+D_{AB}+\frac{1}{4} \gamma_{AB} C^C_DC^D_C+o(r^{-\epsilon})\,, \\
g_{uA} &=&  \frac{1}{2} D_BC^{B}_{A}+ \frac{2}{3}r^{-1}\Big[(\ln r+  \frac{1}{3}) D_BD^{B}_{A}\\
&& +\frac{1}{4} C_{AB}  D_CC^{CB} +N_A\Big]  +o(r^{-1-\varepsilon})\,,\\
 \frac{V}{r}  &=& -\frac{1}{2} R + \frac{2M}{r} + o(r^{-1-\epsilon})
\eea
and $C_A^A = D_A^A = 0$ and indices are raised with the inverse of the unit sphere metric $\gamma_{AB}$ of curvature $R=2$.

The metric of the vacua \eqref{vacua2} (without sources) exactly takes this form where explicitly
\bea
e^{-4\beta} &=& 1 + \frac{C_{AB} C^{AB}}{8r^2},\\
V &=& -r e^{-2\beta}(1+g_{AB}U^A U^B) \quad \Rightarrow \quad M= 0,\\
C_{AB} &=&- (2 D_A D_B -\gamma_{AB} D^2)C,\\
D_{AB} &=& 0,\\
N_A &=&-\frac{3}{32}D_A(C^{BC}C_{BC}) - \frac{1}{4}C_{AB}D_C C^{BC}.
\eea
There is no news $N_{AB} \equiv \p_u C_{AB}$ because $C_{AB}$ is $u$ independent. The fields corresponding to the metric of the vacua with sources \eqref{vacuaRot} can also be readily computed.

Any asymptotically flat metric at past null infinity is the sense of Bondi-van der Burg-Metzner-Sachs can be similarly put in the form
\begin{equation}
 ds^2=e^{2\beta_{(-)}}\frac{V_{(-)}}{r} dv^2 + 2e^{2\beta_{(-)}}dvdr+ g^{(-)}_{AB}(dz^A-U_{(-)}^Adu)(dz^B-U_{(-)}^Bdu).
\end{equation}
We use the same coordinates $z^A$ at $\mathfrak{I}^-$ as in the bulk spacetime and at $\mathfrak{I}^+$: a light ray starting at $(z,\bar z)$ at $\mathfrak{I}^-$ will reach the antipodal point $(-\frac{1}{\bar z},-\frac{1}{z})$ at $\mathfrak{I}^+$. The functions read as
\bea
g^{(-)}_{AB} &=& r^2 \gamma^{(-)}_{AB} + rC^{(-)}_{AB}+D^{(-)}_{AB}+\frac{1}{4} \gamma_{AB} C^{(-)}_{CD}C^{CD}_{(-)}+o(r^{-\epsilon})\,, \\
g^{(-)}_{vA} &=& - \frac{1}{2} D_BC^{B}_{{(-)}A}- \frac{2}{3}r^{-1}\Big[(\ln r+  \frac{1}{3}) D_BD^{B}_{{(-)}A}\\
&& +\frac{1}{4} C^{(-)}_{AB}  D_CC_{(-)}^{CB} +N^{(-)}_A\Big]  +o(r^{-1-\varepsilon})\,,\\
 \frac{V_{(-)}}{r}  &=& -\frac{1}{2} R + r^{-1}2M_{(-)} + o(r^{-1-\epsilon})
\eea
and $C_{{(-)}A}^A = D_{{(-)}A}^A = 0$.  Indices are raised with the inverse of the unit sphere metric $\gamma_{AB}$.

The metric of the vacua \eqref{madv} exactly takes this form where explicitly
\bea
e^{-4\beta^{(-)}} &=& 1 + \frac{C^{(-)}_{AB} C_{(-)}^{AB}}{8r^2},\\
V^{(-)} &=& -r e^{-2\beta^{(-)}}(1+g^{(-)}_{AB}U_{(-)}^A U_{(-)}^B) \quad \Rightarrow \quad M_{(-)}= 0,\\
C^{(-)}_{AB} &=& -(2 D_A D_B -\gamma_{AB} D^2)C^{(-)},\label{defCm}\\
D^{(-)}_{AB} &=& 0,\\
N^{(-)}_A &=&-\frac{3}{32}D_A(C_{(-)}^{BC}C^{(-)}_{BC}) - \frac{1}{4}C^{(-)}_{AB}D^{(-)}_C C_{(-)}^{BC}.
\eea
and
\bea
C(z,\bar z) = C^{(-)}(z,\bar z).\label{match2}
\eea
The definition of $C^{(-)}_{AB}$ in terms of $C^{(-)}$ defines the sign convention for $C^{(-)}$.
 There is no news $N^{(-)}_{AB} \equiv \p_v C^{(-)}_{AB}$ because $C^{(-)}_{AB}$ is $v$ independent. The fields corresponding to the metric of the vacua with sources \eqref{vacuaRot} can be also readily computed.

\subsection{Presymplectic forms and their surface forms}
\label{App:Symp}

The symplectic structure is a key element in the dynamical decription of gravity.  The symplectic structure is a phase space 2-form. It is defined as the integral over a Cauchy slice of a spacetime 3-form that we refer to as the presymplectic form.
The Crnkovic-Witten-Lee-Wald presymplectic form of Einstein gravity \cite{Crnkovic:1986ex,Lee:1990nz}  is given by
\bea
{\boldsymbol \omega}^{LW} [\delta_1 g ,\delta_2 g ; g ] &=& \frac{1}{32\pi G}(d^{3}x)_{\mu}\sqrt{-g } \Big(
\delta_1 g^{\alpha\beta}\nabla^\mu \delta_2 g_{\alpha \beta}+\delta_1 g \nabla^\alpha \delta_2 g^\mu_{\alpha}+\delta_1 g^\mu_{\alpha}\nabla^\alpha \delta_2 g  \nn \\
&&  - \delta_1 g \nabla^\mu \delta_2 g - 2 \delta_1 g_{\alpha\beta}\nabla^\alpha \delta_2 g^{\mu\beta}
- (1 \leftrightarrow 2)
\Big),
\eea
where $\delta_1 g_{\mu\nu}$, $\delta_2 g_{\mu\nu}$ are perturbations around a general asymptotically flat spacetime $g_{\mu\nu}$ and all indices are raised with the inverse metric, $\delta g^{\mu\nu} \equiv g^{\mu \alpha}g^{\nu \beta}\delta g_{\alpha \beta}$.
This integrand is obtained by varying a second time the boundary term $\boldsymbol\Theta [\delta g ; g]$ obtained after a variation of the Einstein-Hilbert Lagrangian as ${\boldsymbol \omega}^{LW} [\delta_1 g ,\delta_2 g ; g ]   = \delta_1 \boldsymbol\Theta[\delta_2 g ; g] - \delta_2 \boldsymbol\Theta[\delta_1 g ; g]$.

The presymplectic form obtained in  \cite{Barnich:2007bf} by acting with a contracting homotopy on Einstein's equations contracted with $\delta g_{\mu\nu}$, is given instead by
\bea
{\boldsymbol \omega}^{BC}  [\delta_1 g ,\delta_2 g ;g ]
&=& \frac{1}{32\pi G}(d^{3}x)_{\mu}\sqrt{-g } \Big(
\delta_1 g^{\alpha\beta}\nabla^\mu \delta_2 g_{\alpha \beta}+\delta_1 g \nabla^\alpha \delta_2 g^\mu_{\alpha}+\delta_1 g^\mu_{\alpha}\nabla^\alpha \delta_2 g \nn \\
&&  - \delta_1 g \nabla^\mu \delta_2 g - \delta_1 g_{\alpha\beta}\nabla^\alpha \delta_2 g^{\mu\beta} - \delta_1 g^{\mu\alpha}\nabla^\beta \delta_2 g_{\alpha \beta}
- (1 \leftrightarrow 2)
\Big).
\eea
It was called the invariant presymplectic form in \cite{Barnich:2007bf} in reference to the fact that it only depends upon the equations of motion of the action instead of the action itself. 

The difference between the two presymplectic structures is a boundary term, ${\boldsymbol \omega}^{BC} - {\boldsymbol \omega}^{LW} = d {\boldsymbol E}$ which is given by
\bea
{\boldsymbol \omega}^{BC}  [\delta_1 g ,\delta_2 g ;g] - {\boldsymbol \omega}^{LW}  [\delta_1 g ,\delta_2 g ;g]  = \frac{1}{32\pi G}(d^{3}x)_{\mu}\sqrt{-g } \nabla_\nu \Big( \delta_1 g^\nu_{\; \beta} \delta_2 g^{\mu\beta} - (\mu \rightarrow \nu)\Big).
\eea
or
\bea
\boldsymbol E[\delta_1 g , \delta_2 g ; g]  = - \frac{\sqrt{-g}}{32 \pi G} (d^{2}x)_{\mu \nu }  \Big( (\delta_1 g)^{\mu \rho} (\delta_2 g)_\rho^{\;\; \nu} - (\mu \rightarrow \nu)\Big).\label{defE}
\eea

The surface forms which are canonically associated with these presymplectic forms are defined from the \emph{off-shell} equality
\bea
{\boldsymbol \omega}^{BC}[\delta g, \mathcal L_\xi g  ; g ] &=& d \boldsymbol k^{AD}_\xi [ \delta g ; g ]+ (EOM)+ \delta (EOM) ,\\
{\boldsymbol \omega}^{LW}[\delta g, \mathcal L_\xi g ; g ] &=& d \boldsymbol k^{IW}_\xi [ \delta g ; g ]+ (EOM)+ \delta (EOM)\label{defk}
\eea
where $(EOM)$ are terms proportional to the equations of motion and $\delta (EOM)$ are terms proportional to the linearized equations of motion.

We adopt the standard definition for the homotopy operator $I_\xi$ which obeys $d I_\xi + I_\xi d = 1$ upon acting on an expression linear in $\xi$ and its derivatives \cite{Barnich:2007bf,Compere:2007az}. The surface forms associated with these presymplectic forms are then defined as
\bea
\boldsymbol k^{AD}_\xi [ \delta g ; g ] &=& I_\xi {\boldsymbol \omega}^{BC}[ \delta g, \mathcal L_\xi g ; g ] ,\\
 \boldsymbol k^{IW}_\xi [ \delta g ; g ] &=&  I_\xi {\boldsymbol \omega}^{LW}[ \delta g, \mathcal L_\xi g ; g ] .
\eea
Note that the definition \eqref{defk} does not fix the codimension 3 boundary terms in the definition of the surface forms. The procedure of using an homotopy only gives a definite prescription.

The surface charge form $\boldsymbol k^{AD}_\xi [ \delta g ; g ]$ is exactly given by the Abbott-Deser expression \cite{Abbott:1981ff} or equivalently the Barnich-Brandt expression \cite{Barnich:2001jy} for the surface charge for a linear perturbation $\delta g_{\mu\nu}$ around a solution $g_{\mu\nu}$
\bea
\boldsymbol k^{AD}_\xi [\delta g ; g ]
&=& \frac{1}{16\pi  G}(d^{2}x)_{\mu\nu}\sqrt{-g } \Big( \xi^\nu (D^\mu \delta g - D_\sigma \delta g^{\sigma \mu}) +\xi_\sigma D^\nu \delta g^{\sigma \mu} \nonumber \\
& &+\frac{1}{2} \delta g D^\nu \xi^\mu +\frac{1}{2}\delta g^{\mu \sigma}D_\sigma \xi^\nu +\frac{1}{2} \delta g^{\nu \sigma}D^\mu \xi_\sigma - (\mu \leftrightarrow \nu) \Big) \\
&=&\boldsymbol  k^{IW}_\xi [\delta g ; g ]  - \boldsymbol E[\mathcal L_\xi g, \delta g ; g].
\label{ADcharge}
\eea
In the last equivalent form, $\boldsymbol E$ is defined in \eqref{defE} and $\boldsymbol k^{IW}_\xi [\delta g ; g ] $ is the Iyer-Wald surface form which is canonically associated with the Lee-Wald presymplectic structure. It is given by
\bea
\boldsymbol k^{IW}_\xi [\delta g ; g ] &=& - \delta \boldsymbol  k^K_\xi + \boldsymbol  k^K_{\delta \xi}+  \frac{1}{8\pi G}(d^{2}x)_{\mu\nu}\sqrt{-g } \Big( \xi^\nu (D^\mu \delta g - D_\sigma \delta g^{\sigma \mu})+D^\mu \delta \xi^\nu \Big) \label{kk1},
\eea
where
\bea
\boldsymbol k^K_\xi[g] =  \frac{1}{16\pi G}(d^{2}x)_{\mu\nu}\sqrt{-g } \Big( D^\mu \xi^\nu - D^\nu \xi^\mu \Big)\, ,
\eea
is the Komar surface form.

It is important to note that there is an ambiguity in the definition of the presymplectic potential \eqref{Theta} under the addition of a total derivative,
\bea
\Theta^\mu[ \delta g ; g] \rightarrow \Theta^\mu[ \delta g ; g] + \p_\nu \Theta_{amb}^{\mu\nu}[ \delta g ; g]\label{amb}
\eea
where $\Theta_{amb}^{\mu\nu} = \Theta_{amb}^{[\mu\nu]}$. The ambiguity \eqref{amb} leads to the following ambiguity in the definition of the Lee-Wald presymplectic form and Iyer-Wald surface charge,
\bea
 {\boldsymbol \omega}_{LW}[\delta_1 g , \delta_2 g ; g] &\rightarrow &  {\boldsymbol \omega}_{LW}[\delta_1 g , \delta_2 g ; g]  +d \left(  \delta_1 {\boldsymbol \Theta}_{amb}[ \delta_2 g ;g]    -  \delta_2  {\boldsymbol \Theta}_{amb}  [\delta_1 g ; g]  \right), \\
  {\boldsymbol k}^{IW}_\xi [ \delta g ; g] & \rightarrow & {\boldsymbol k}^{IW}_\xi [ \delta g ; g] + \mathcal L_\xi  {\boldsymbol \Theta}_{amb}  [\delta g ; g] -  \delta  {\boldsymbol \Theta}_{amb}  [\mathcal L_\xi g ; g] .
 \eea

\subsection{Proof of the charge formula \eqref{QRO}}
\label{proof1}

The proof of the equivalence of the charges \eqref{QR} and \eqref{QRO} for rotations and Lorentz generators goes as follows. Let us first obtain some useful properties of generic superrotation functions $R^A=(R(z),\bar R(\bar z))$:
\bea
(D^2+1)R^A = 0,\qquad (D^2+2)D_A R^A = 0, \qquad D_A R_B + D_B R_A = \gamma_{AB} D_C R^C.\label{propR}
\eea
Also note that for any vector $V^A$ and scalar $S$ on the sphere we have
\bea
[D^2,D_A]V^A = -D_A V^A,\qquad [D^2,D_A]S = D_A S.
\eea

Let us start massaging the expression for the charges \eqref{QR}. Since the global conformal Killing vectors are globally defined, we can freely integrate by parts and drop the boundary terms. After one integration by parts, using the definition of $C_{AB}$ and the property \eqref{propC} we obtain
\bea
\mathcal Q^S_{\chi_R}&=&\frac{1}{16G} \oint_S d^2\Omega (D_AR^A)(2D_ED_F-\gamma_{EF}D^2)C D^ED^FC  +2R^A C_{AB} D^B(D^2+2)C.\nn
\eea
We then integrate by parts the $D^B$ in the final term and we get two kinds of new terms. For those proportional to $D^B R^A$ we can use the conformal Killing equation to recognize that it is proportional to $\gamma^{AB}$ but $C_{AB}$ is traceless so these terms are zero. For those proportional to $D^BC_{AB}$ we use again property \eqref{propC} and perform an integration by parts. We then obtain
\bea
\mathcal Q^S_{\chi_R}&=&\frac{1}{16G} \oint_S d^2\Omega 2 D_AR^A D_ED_FC D^ED^FC- D_AR^A (D^2C)^2- D_AR^A((D^2+2)C)^2.\nn
\eea
We then perform an additional integration by parts and we distribute the last term to obtain
\bea
\mathcal Q^S_{\chi_R} &=&\frac{1}{8G}  \oint_S d^2\Omega(- D_AD_ER^E D_BC D^AD^BC- D_AR^A D_BCD^2D^BC- D_AR^A(D^2C)^2\nn\\
&&-2 D_AR^ACD^2C- 2D_AR^AC^2).\label{eq:Q6}
\eea
The first term in the parenthesis can be written as $ - \frac{1}{2} D_AD_ER^E D^A( D_BC D^BC)$ which can then be integrated by parts as $+ \frac{1}{2} D^2 D_ER^E ( D_BC D^BC)$. The second term can be written as
\bea
- D_AR^A D_BCD^2D^BC &=& - D_AR^A D_BC (D^B D^2 C +D^B C) \nn \\
&=& D^B D_A R^A D_B C D^2 C + D_A R^A (D^2 C)^2 - D_AR^A D_BC D^B C.\nn
\eea
The third term in \eqref{eq:Q6} will then be canceled. The fourth term can be written as
\bea
-2 D_AR^ACD^2C &=& 2 D_E D_A R^A C D^E C +2 D_A R^A D_E C D^E C\nn \\
&=&   D_E D_A R^A D^E (C^2) +2 D_A R^A D_E C D^E C\nn\\
&=& -D^2 D_A R^A C^2 +2 D_A R^A D_E C D^E C\nn\\
&=& 2 D_A R^A C^2 +2 D_A R^A D_E C D^E C.
\eea
The fifth term in \eqref{eq:Q6} will then be canceled. Regrouping all the remaining terms we have
\bea
\mathcal Q^S_{\chi_R} &=&\frac{1}{8G} \oint_S d^2\Omega  \frac{1}{2} D^2 D_ER^E ( D_BC D^BC)+D_B D_E R^E D^B C D^2 C + D_AR^A D_BC D^B C.
\nn
\eea
Let us massage the second term in the parenthesis
\bea
D_B D_E R^E D^B C D^2 C &=& - D_A D_B D_E R^E D^B C D^A C -   D_B D_E R^E D_A D^B C D^A C \nn\\
&=& - D_A D_B D_E R^E D^B C D^A C -   D_B D_E R^E D^B D_A C D^A C\nn\\
&=& - D_A D_B D_E R^E D^B C D^A C -  \frac{1}{2} D_B D_E R^E D^B (D_A C D^A C)\nn\\
&=& - D_A D_B D_E R^E D^B C D^A C +  \frac{1}{2} D^2 D_E R^E  D_A C D^A C.
\eea
We then obtain
\bea
\mathcal Q^S_{\chi_R} &=&\frac{1}{8G} \oint_S d^2\Omega  D^2 D_ER^E ( D_BC D^BC)- D_A D_B D_E R^E D^B C D^A C + D_AR^A D_BC D^B C.
\nn
\eea
We finally use the second property \eqref{propR} to obtain the final result \eqref{QRO}.


\begin{thebibliography}{10}

\bibitem{Bondi:1962px}
H.~Bondi, M.~G.~J. van~der Burg, and A.~W.~K. Metzner, ``{Gravitational waves
  in general relativity. 7. Waves from axisymmetric isolated systems},'' {\em
  Proc. Roy. Soc. Lond.} {\bf A269} (1962)
21--52.

\bibitem{Sachs:1962wk}
R.~K. Sachs, ``{Gravitational waves in general relativity. 8. Waves in
  asymptotically flat space-times},'' {\em Proc. Roy. Soc. Lond.} {\bf A270}
  (1962)
103--126.

\bibitem{ABR}
L.~B. A.~Ashtekar and O.~Reula, ``{The Covariant Phase Space Of Asymptotically
  Flat Gravitational Fields},'' {\em Analysis, Geometry and Mechanics: 200
  Years After Lagrange, edited by M. Francaviglia and D. Holm (North-Holland,
  Amsterdam, 1991)}.

\bibitem{Strominger:2013jfa}
A.~Strominger, ``{On BMS Invariance of Gravitational Scattering},'' {\em JHEP}
  {\bf 07} (2014) 152,
\href{http://www.arXiv.org/abs/1312.2229}{{\tt 1312.2229}}.

\bibitem{He:2014laa}
T.~He, V.~Lysov, P.~Mitra, and A.~Strominger, ``{BMS supertranslations and
  Weinberg's soft graviton theorem},'' {\em JHEP} {\bf 05} (2015) 151,
\href{http://www.arXiv.org/abs/1401.7026}{{\tt 1401.7026}}.

\bibitem{Strominger:2014pwa}
A.~Strominger and A.~Zhiboedov, ``{Gravitational Memory, BMS Supertranslations
  and Soft Theorems},'' {\em JHEP} {\bf 01} (2016) 086,
\href{http://www.arXiv.org/abs/1411.5745}{{\tt 1411.5745}}.

\bibitem{Brown:1986nw}
J.~D. Brown and M.~Henneaux, ``{Central Charges in the Canonical Realization of
  Asymptotic Symmetries: An Example from Three-Dimensional Gravity},'' {\em
  Commun. Math. Phys.} {\bf 104} (1986)
207--226.

\bibitem{Fefferman:1985aa}
C.~Fefferman and C.~Robin~Graham, ``{Conformal Invariants},'' {\em in Elie
  Cartan et les Math\'ematiques d'aujourd'hui (Ast\'erisque)} (1985) 95.

\bibitem{Banados:1998gg}
M.~Banados, ``{Three-dimensional quantum geometry and black holes},''
  \href{http://www.arXiv.org/abs/hep-th/9901148}{{\tt hep-th/9901148}}.
[AIP Conf. Proc.484,147(1999)].

\bibitem{Rooman:2000ei}
M.~Rooman and P.~Spindel, ``{Uniqueness of the asymptotic AdS(3) geometry},''
  {\em Class. Quant. Grav.} {\bf 18} (2001) 2117--2124,
\href{http://www.arXiv.org/abs/gr-qc/0011005}{{\tt gr-qc/0011005}}.

\bibitem{Barnich:2010eb}
G.~Barnich and C.~Troessaert, ``{Aspects of the BMS/CFT correspondence},'' {\em
  JHEP} {\bf 05} (2010) 062,
\href{http://www.arXiv.org/abs/1001.1541}{{\tt 1001.1541}}.

\bibitem{Barnich:2012rz}
G.~Barnich, A.~Gomberoff, and H.~A. González, ``{Three-dimensional
  Bondi-Metzner-Sachs invariant two-dimensional field theories as the flat
  limit of Liouville theory},'' {\em Phys. Rev.} {\bf D87} (2013), no.~12,
  124032,
\href{http://www.arXiv.org/abs/1210.0731}{{\tt 1210.0731}}.

\bibitem{Barnich:2013yka}
G.~Barnich and H.~A. Gonzalez, ``{Dual dynamics of three dimensional
  asymptotically flat Einstein gravity at null infinity},'' {\em JHEP} {\bf 05}
  (2013) 016,
\href{http://www.arXiv.org/abs/1303.1075}{{\tt 1303.1075}}.

\bibitem{Garbarz:2014kaa}
A.~Garbarz and M.~Leston, ``{Classification of Boundary Gravitons in AdS$_3$
  Gravity},'' {\em JHEP} {\bf 05} (2014) 141,
\href{http://www.arXiv.org/abs/1403.3367}{{\tt 1403.3367}}.

\bibitem{Barnich:2014kra}
G.~Barnich and B.~Oblak, ``{Notes on the BMS group in three dimensions: I.
  Induced representations},'' {\em JHEP} {\bf 06} (2014) 129,
\href{http://www.arXiv.org/abs/1403.5803}{{\tt 1403.5803}}.

\bibitem{Barnich:2015uva}
G.~Barnich and B.~Oblak, ``{Notes on the BMS group in three dimensions: II.
  Coadjoint representation},'' {\em JHEP} {\bf 03} (2015) 033,
\href{http://www.arXiv.org/abs/1502.00010}{{\tt 1502.00010}}.

\bibitem{Compere:2015knw}
G.~Comp\`ere, P.-J. Mao, A.~Seraj, and S.~Sheikh-Jabbari, ``{Symplectic and
  Killing Symmetries of AdS$_3$ Gravity: Holographic vs Boundary Gravitons},''
  {\em JHEP} {\bf 01} (2016) 080,
\href{http://www.arXiv.org/abs/1511.06079}{{\tt 1511.06079}}.

\bibitem{Banados:1992wn}
M.~Banados, C.~Teitelboim, and J.~Zanelli, ``{The Black hole in
  three-dimensional space-time},'' {\em Phys. Rev. Lett.} {\bf 69} (1992)
  1849--1851,
\href{http://www.arXiv.org/abs/hep-th/9204099}{{\tt hep-th/9204099}}.

\bibitem{Deser:1984dr}
S.~Deser and R.~Jackiw, ``Three-dimensional cosmological gravity: Dynamics of
  constant curvature,'' {\em Annals Phys.} {\bf 153} (1984)
405--416.

\bibitem{Barnich:2011mi}
G.~Barnich and C.~Troessaert, ``{BMS charge algebra},'' {\em JHEP} {\bf 12}
  (2011) 105,
\href{http://www.arXiv.org/abs/1106.0213}{{\tt 1106.0213}}.

\bibitem{Barnich:2013oba}
G.~Barnich and P.-H. Lambert, ``{A note on the Newman-Unti group and the BMS
  charge algebra in terms of Newman-Penrose coefficients},'' {\em J. Phys.
  Conf. Ser.} {\bf 410} (2013)
012142.

\bibitem{Compere:2014cna}
G.~Comp\`ere, L.~Donnay, P.-H. Lambert, and W.~Schulgin, ``{Liouville theory
  beyond the cosmological horizon},'' {\em JHEP} {\bf 03} (2015) 158,
\href{http://www.arXiv.org/abs/1411.7873}{{\tt 1411.7873}}.

\bibitem{Compere:2015bca}
G.~Comp\`ere, K.~Hajian, A.~Seraj, and M.~M. Sheikh-Jabbari, ``{Wiggling Throat
  of Extremal Black Holes},'' {\em JHEP} {\bf 10} (2015) 093,
\href{http://www.arXiv.org/abs/1506.07181}{{\tt 1506.07181}}.

\bibitem{Flanagan:2015pxa}
E.~E. Flanagan and D.~A. Nichols, ``{Conserved charges of the extended
  Bondi-Metzner-Sachs algebra},''
\href{http://www.arXiv.org/abs/1510.03386}{{\tt 1510.03386}}.

\bibitem{Hawking:2016msc}
S.~W. Hawking, M.~J. Perry, and A.~Strominger, ``{Soft Hair on Black Holes},''
  {\em Phys.Rev.Lett.} {\bf 116} (2016), no.~23, 231301,
\href{http://www.arXiv.org/abs/1601.00921}{{\tt 1601.00921}}.

\bibitem{Barnich:2009se}
G.~Barnich and C.~Troessaert, ``{Symmetries of asymptotically flat 4
  dimensional spacetimes at null infinity revisited},'' {\em Phys. Rev. Lett.}
  {\bf 105} (2010) 111103,
\href{http://www.arXiv.org/abs/0909.2617}{{\tt 0909.2617}}.

\bibitem{Banks:2003vp}
T.~Banks, ``{A Critique of pure string theory: Heterodox opinions of diverse
  dimensions},''
\href{http://www.arXiv.org/abs/hep-th/0306074}{{\tt hep-th/0306074}}.

\bibitem{Kapec:2014opa}
D.~Kapec, V.~Lysov, S.~Pasterski, and A.~Strominger, ``{Semiclassical Virasoro
  symmetry of the quantum gravity $ \mathcal{S}$-matrix},'' {\em JHEP} {\bf 08}
  (2014) 058,
\href{http://www.arXiv.org/abs/1406.3312}{{\tt 1406.3312}}.

\bibitem{Pasterski:2015tva}
S.~Pasterski, A.~Strominger, and A.~Zhiboedov, ``{New Gravitational
  Memories},''
\href{http://www.arXiv.org/abs/1502.06120}{{\tt 1502.06120}}.

\bibitem{Troessaert:2015nia}
C.~Troessaert, ``{Hamiltonian surface charges using external sources},'' {\em
  J. Math. Phys.} {\bf 57} (2016), no.~5, 053507,
\href{http://www.arXiv.org/abs/1509.09094}{{\tt 1509.09094}}.

\bibitem{Compere:2016hzt}
G.~Comp\`ere and J.~Long, ``{Classical static final state of collapse with
  supertranslation memory},''
\href{http://www.arXiv.org/abs/1602.05197}{{\tt 1602.05197}}.

\bibitem{Regge:1974zd}
T.~Regge and C.~Teitelboim, ``{Role of Surface Integrals in the Hamiltonian
  Formulation of General Relativity},'' {\em Annals Phys.} {\bf 88} (1974)
286.

\bibitem{Barnich:2011ct}
G.~Barnich and C.~Troessaert, ``{Supertranslations call for superrotations},''
  {\em PoS} (2010) 010, \href{http://www.arXiv.org/abs/1102.4632}{{\tt
  1102.4632}}.
[Ann. U. Craiova Phys.21,S11(2011)].

\bibitem{Balog:1997zz}
J.~Balog, L.~Feher, and L.~Palla, ``{Coadjoint orbits of the Virasoro algebra
  and the global Liouville equation},'' {\em Int. J. Mod. Phys.} {\bf A13}
  (1998) 315--362,
\href{http://www.arXiv.org/abs/hep-th/9703045}{{\tt hep-th/9703045}}.

\bibitem{Barnich:2001jy}
G.~Barnich and F.~Brandt, ``{Covariant theory of asymptotic symmetries,
  conservation laws and central charges},'' {\em Nucl. Phys.} {\bf B633} (2002)
  3--82,
\href{http://www.arXiv.org/abs/hep-th/0111246}{{\tt hep-th/0111246}}.

\bibitem{Christodoulou:1993uv}
D.~Christodoulou and S.~Klainerman, ``{The Global nonlinear stability of the
  Minkowski space},'' {\em Princeton University Press, Princeton}
(1993).

\bibitem{Balasubramanian:1999re}
V.~Balasubramanian and P.~Kraus, ``{A Stress tensor for Anti-de Sitter
  gravity},'' {\em Commun. Math. Phys.} {\bf 208} (1999) 413--428,
\href{http://www.arXiv.org/abs/hep-th/9902121}{{\tt hep-th/9902121}}.

\bibitem{Iyer:1994ys}
V.~Iyer and R.~M. Wald, ``{Some properties of Noether charge and a proposal for
  dynamical black hole entropy},'' {\em Phys. Rev.} {\bf D50} (1994) 846--864,
\href{http://www.arXiv.org/abs/gr-qc/9403028}{{\tt gr-qc/9403028}}.

\bibitem{Barnich:1995ap}
G.~Barnich, F.~Brandt, and M.~Henneaux, ``{Local BRST cohomology in Einstein
  Yang-Mills theory},'' {\em Nucl. Phys.} {\bf B455} (1995) 357--408,
\href{http://www.arXiv.org/abs/hep-th/9505173}{{\tt hep-th/9505173}}.

\bibitem{Barnich:2000zw}
G.~Barnich, F.~Brandt, and M.~Henneaux, ``{Local BRST cohomology in gauge
  theories},'' {\em Phys. Rept.} {\bf 338} (2000) 439--569,
\href{http://www.arXiv.org/abs/hep-th/0002245}{{\tt hep-th/0002245}}.

\bibitem{Barnich:2007bf}
G.~Barnich and G.~Comp\`ere, ``{Surface charge algebra in gauge theories and
  thermodynamic integrability},'' {\em J. Math. Phys.} {\bf 49} (2008) 042901,
\href{http://www.arXiv.org/abs/0708.2378}{{\tt 0708.2378}}.

\bibitem{Dickenstein:1999gb}
A.~Dickenstein, M.~S. Iriondo, and T.~A. Rojas, ``{Integrating singular
  functions on the sphere},'' {\em J. Math. Phys.} {\bf 38} (1997) 5361--5370,
\href{http://www.arXiv.org/abs/gr-qc/9902013}{{\tt gr-qc/9902013}}.

\bibitem{Witten:1998qj}
E.~Witten, ``{Anti-de Sitter space and holography},'' {\em Adv. Theor. Math.
  Phys.} {\bf 2} (1998) 253--291,
\href{http://www.arXiv.org/abs/hep-th/9802150}{{\tt hep-th/9802150}}.

\bibitem{Barnich:2003xg}
G.~Barnich, ``{Boundary charges in gauge theories: Using Stokes theorem in the
  bulk},'' {\em Class. Quant. Grav.} {\bf 20} (2003) 3685--3698,
\href{http://www.arXiv.org/abs/hep-th/0301039}{{\tt hep-th/0301039}}.

\bibitem{Deser:1983tn}
S.~Deser, R.~Jackiw, and G.~'t~Hooft, ``{Three-Dimensional Einstein Gravity:
  Dynamics of Flat Space},'' {\em Annals Phys.} {\bf 152} (1984)
220.

\bibitem{Ashtekar:1996cd}
A.~Ashtekar, J.~Bicak, and B.~G. Schmidt, ``{Asymptotic structure of symmetry
  reduced general relativity},'' {\em Phys. Rev.} {\bf D55} (1997) 669--686,
\href{http://www.arXiv.org/abs/gr-qc/9608042}{{\tt gr-qc/9608042}}.

\bibitem{Barnich:2006av}
G.~Barnich and G.~Comp\`ere, ``{Classical central extension for asymptotic
  symmetries at null infinity in three spacetime dimensions},'' {\em Class.
  Quant. Grav.} {\bf 24} (2007) F15--F23,
\href{http://www.arXiv.org/abs/gr-qc/0610130}{{\tt gr-qc/0610130}}.

\bibitem{Vilenkin:2000jqa}
A.~Vilenkin and E.~P.~S. Shellard, {\em {Cosmic Strings and Other Topological
  Defects}}.
\newblock Cambridge University Press,
2000.
\newblock

\bibitem{Witten:1985fp}
E.~Witten, ``{Cosmic Superstrings},'' {\em Phys. Lett.} {\bf B153} (1985)
243.

\bibitem{Copeland:2003bj}
E.~J. Copeland, R.~C. Myers, and J.~Polchinski, ``{Cosmic F and D strings},''
  {\em JHEP} {\bf 06} (2004) 013,
\href{http://www.arXiv.org/abs/hep-th/0312067}{{\tt hep-th/0312067}}.

\bibitem{Hossenfelder:2014hha}
S.~Hossenfelder, ``{Theory and Phenomenology of Spacetime Defects},'' {\em Adv.
  High Energy Phys.} {\bf 2014} (2014) 950672,
\href{http://www.arXiv.org/abs/1401.0276}{{\tt 1401.0276}}.

\bibitem{Bousso:1999xy}
R.~Bousso, ``{A Covariant entropy conjecture},'' {\em JHEP} {\bf 07} (1999)
  004,
\href{http://www.arXiv.org/abs/hep-th/9905177}{{\tt hep-th/9905177}}.

\bibitem{Oblak:2015sea}
B.~Oblak, ``{Characters of the BMS Group in Three Dimensions},'' {\em Commun.
  Math. Phys.} {\bf 340} (2015), no.~1, 413--432,
\href{http://www.arXiv.org/abs/1502.03108}{{\tt 1502.03108}}.

\bibitem{Barnich:2015mui}
G.~Barnich, H.~A. Gonzalez, A.~Maloney, and B.~Oblak, ``{One-loop partition
  function of three-dimensional flat gravity},'' {\em JHEP} {\bf 04} (2015)
  178,
\href{http://www.arXiv.org/abs/1502.06185}{{\tt 1502.06185}}.

\bibitem{Garbarz:2015lua}
A.~Garbarz and M.~Leston, ``{Quantization of BMS$_3$ orbits: a perturbative
  approach},'' {\em Nucl.Phys.B} {\bf 906} (2016) 133--146,
\href{http://www.arXiv.org/abs/1507.00339}{{\tt 1507.00339}}.

\bibitem{Banerjee:2015kcx}
N.~Banerjee, D.~P. Jatkar, S.~Mukhi, and T.~Neogi, ``{Free-field realisations
  of BMS$_3$ and super-BMS$_3$ algebras},'' {\em JHEP} {\bf 06} (2016) 024,
\href{http://www.arXiv.org/abs/1512.06240}{{\tt 1512.06240}}.

\bibitem{Zeldovich:1974aa}
B.~Zeldovich and A.~G. Polnarev., ``Radiation of gravitational waves by a
  cluster of superdense stars,'' {\em Ya. Sov.Astron.Lett.} {\bf 18} (1974) 17.

\bibitem{Christodoulou:1991cr}
D.~Christodoulou, ``{Nonlinear nature of gravitation and gravitational wave
  experiments},'' {\em Phys. Rev. Lett.} {\bf 67} (1991)
1486--1489.

\bibitem{Crnkovic:1986ex}
C.~Crnkovic and E.~Witten, ``Covariant description of canonical formalism in
  geometrical theories, 1986, \textit{Print-86-1309 (Princeton)},''.

\bibitem{Lee:1990nz}
J.~Lee and R.~M. Wald, ``{Local symmetries and constraints},'' {\em J. Math.
  Phys.} {\bf 31} (1990)
725--743.

\bibitem{Compere:2007az}
G.~Comp\`ere, {\em {Symmetries and conservation laws in Lagrangian gauge
  theories with applications to the mechanics of black holes and to gravity in
  three dimensions}}.
\newblock PhD thesis, Vrije U., Brussels, 2007.
\newblock
\href{http://www.arXiv.org/abs/0708.3153}{{\tt 0708.3153}}.
\newblock

\bibitem{Abbott:1981ff}
L.~F. Abbott and S.~Deser, ``{Stability of Gravity with a Cosmological
  Constant},'' {\em Nucl. Phys.} {\bf B195} (1982)
76.

\end{thebibliography}

\providecommand{\href}[2]{#2}\begingroup\raggedright\endgroup

\end{document}